\documentclass[fleqn,usenatbib]{mnras}

\usepackage{newtxtext,newtxmath}

\usepackage[T1]{fontenc}

\DeclareRobustCommand{\VAN}[3]{#2}
\let\VANthebibliography\thebibliography
\def\thebibliography{\DeclareRobustCommand{\VAN}[3]{##3}\VANthebibliography}

\usepackage{graphicx}	% Including figure files
\usepackage{amsmath}	% Advanced maths commands
\usepackage{amssymb}	% Extra maths symbols
\usepackage{booktabs}
\usepackage{float}
\usepackage{multicol}

\newcommand{\arf}{\color{black}}

\newcommand{\ASL}{\color{black}}

\newcommand{\oIII}{\mbox{O~{\sc iii}}}

\newcommand{\muv}{\ifmmode M_{\rm UV} \else $M_{\rm UV}$\fi}

\newcommand{\ebv}{\ifmmode E_{\rm B-V} \else $E_{\rm B-V}$\fi}

\newcommand{\sigSFR}{\ifmmode \Sigma_{\rm SFR} \else $\Sigma_{\rm SFR}$\fi}

\newcommand{\mdyn}{\ifmmode M_{\rm dyn} \else $M_{\rm dyn}$\fi}
\newcommand{\mstar}{\ifmmode M_{\star} \else $M_{\star}$\fi}
\newcommand{\msun}{\ifmmode M_{\odot} \else $M_{\odot}$\fi}

\newcommand{\vflow}{\ifmmode v_{\rm flow} \else $v_{\rm flow}$\fi}
\newcommand{\mflow}{\ifmmode \dot{M}_{\rm flow} \else $\dot{M}_{\rm flow}$\fi}

\title[Feedback and dynamical masses at high-$z$]{Feedback and dynamical masses in high-$z$ galaxies: the advent of high-resolution NIRSpec spectroscopy}

\author[A. Saldana-Lopez et al.]{
A. Saldana-Lopez,$^{1}$ \thanks{E-mail: alberto.saldana-lopez@astro.su.se}
J. Chisholm,$^{2}$
S. Gazagnes,$^{2}$
R. Endsley,$^{2}$
M. J. Hayes,$^{1}$
\newauthor
D. A. Berg,$^{2}$
S. L. Finkelstein,$^{2}$ 
S. R. Flury,$^{3}$
N. G. Guseva,$^{4}$
A. Henry,$^{5}$
Y. I. Izotov,$^{4}$
\newauthor
E. Lambrides,$^{6}$
R. Marques-Chaves,$^{7}$
C. T. Richardson,$^{8}$\\
$^{1}$ Department of Astronomy, Oskar Klein Centre, Stockholm University, 106 91 Stockholm, Sweden\\
$^{2}$ Department of Astronomy, The University of Texas at Austin, 2515 Speedway, Stop C1400, Austin, TX 78712-1205, USA\\
$^{3}$ Institute for Astronomy, University of Edinburgh, Royal Observatory, Edinburgh, EH9 3HJ, UK\\
$^{4}$ Bogolyubov Institute for Theoretical Physics, National Academy of Sciences of Ukraine, 14-b Metrolohichna str., Kyiv, 03143, Ukraine\\
$^{5}$ Space Telescope Science Institute, 3700 San Martin Drive Baltimore, MD 21218, USA\\
$^{6}$ Astrophysics Science Division, Code 662, NASA Goddard Space Flight Center, 8800 Greenbelt Rd, Greenbelt, MD 20771, USA\\
$^{7}$ Department of Astronomy, University of Geneva, 51 Chemin Pegasi, 1290 Versoix, Switzerland\\
$^{8}$ Department of Physics \& Astronomy, Elon University, 100 Campus Drive, Elon, NC 27244
}

\date{Accepted 2025 September 26. Received 2025 September 9; in original form 2025 January 28}

\pubyear{2025}

\begin{document}
\label{firstpage}
\pagerange{\pageref{firstpage}--\pageref{lastpage}}
\maketitle

% Abstract of the paper
\begin{abstract}
Stellar feedback is an essential step in the baryon cycle of galaxies, but it remains unconstrained beyond Cosmic Noon. We study the dynamical mass and gas-flow properties of a sample of 16 sub-$L^{\star}$ star-forming galaxies at $4\leq z\leq7.6$, using high-resolution JWST/NIRSpec observations. From the velocity dispersion of the (resolved) emission lines ($\sigma_{\rm gas}{\rm~(km~s^{-1})}\simeq38-96$) and the galaxy size ($r_e=400-960~$pc), we estimate dynamical masses of $\log M_{\rm dyn}/M_{\odot}=9.25-10.25$. Stellar-to-dynamical mass ratios are low ($\log M_{\star}/M_{\rm dyn}\in[-0.5,-2]$) and decrease with increasing SFR surface-density ($\Sigma_{\rm SFR}$). We estimate gas surface-densities assuming a star-formation law, but the gas masses do not balance the baryon-to-dynamical mass ratios, requiring a lower star-formation efficiency. Evidence of ionized outflows is found in five galaxies, based on broad components reproducing the emission-line wings. We only observe outflows from galaxies undergoing recent bursts of star formation ${\rm SFR_{10}/SFR_{100}\geq1}$, with elevated $\Sigma_{\rm SFR}$ and low $M_{\star}/M_{\rm dyn}$. This links high gas surface-densities to increased outflow incidence and lower $M_{\star}/M_{\rm dyn}$. With moderate outflow velocities ($v_{\rm flow}{\rm~(km~s^{-1})}=150-250$) and mass outflow rates ($\dot{M}_{\rm flow}/{\rm M_{\odot} yr^{-1}}=0.2-5$), these high-redshift galaxies appear more efficient at removing baryons than low-redshift galaxies with similar $M_{\star}$, showing mass loading-factors of $\dot{M}_{\rm flow}/{\rm SFR}=0.04-0.4$. For their dynamical mass, outflow velocities exceed the escape velocities, meaning they may eventually enrich the Circumgalactic Medium. 
\end{abstract}

% Select between one and six entries from the list of approved keywords.
% Don't make up new ones.

%%%%%%%%%%%%%%%%%%%%%%%%%%%%%%%%%%%%%%%%%%%%%%%%%%

%%%%%%%%%%%%%%%%% BODY OF PAPER %%%%%%%%%%%%%%%%%%

\defcitealias{R16}{R16}

\newpage
\section{Introduction}\label{sec:intro}
{\ASL Gas in galaxies collapses under gravity in the depths of potential wells to forge stellar mass.} However, galaxies are inefficient at converting gas into stars \citep[e.g., see][]{WhiteFrenk1991, Vogelsberger2013, Wechsler2018review}. Abundance matching suggests that the stellar-to-halo mass ratio is below the constant $\simeq 20\%$ of the cosmic baryon abundance predicted by the $\Lambda$CDM formalism, reaching its maximum at halo masses of $\log M_{h}/\msun \simeq 12$, and declining on both sides of this mass \citep[e.g.,][]{Conroy2009, Behroozi2013b}. {\ASL This decreased $\mstar/M_{h}$ ratio can be explained if significant amounts of baryons are either prevented from forming stars or removed from the star-forming regions, via the so-called \emph{feedback} mechanisms \citep[e.g.,][]{Hernquist2003, Hayward2017}, while non-baryons remain within the halos.} 

Different feedback modes will operate at the high or low mass ends of the halo mass function \citep[e.g.,][]{DekelSilk1986, Fabian2012}. {\ASL In the high mass regime ($\log M_{h}/\msun > 12$), thermal feedback from active galactic nuclei (AGN) reduces the ability of the gas to cool to further accrete and form stars \citep[e.g.,][]{ReesOstriker1977}. Together with the mechanical and radiative feedback in the form of jets and energetic outflows \citep[e.g.,][]{HeckmanBest2014}, both processes may regulate the star formation in such high-mass systems \citep[e.g.,][]{VillarMartin2011, Leung2019, Revalski2021, RuschelDutra2021, Flury2023}, resulting in a co-evolution between the mass assembly of the host and the AGN activity \citep[e.g.,][]{Harrison2014, Fiore2017}.} 

In low-mass haloes, feedback from massive stars dominates \citep[e.g.,][]{Veilleux2005, Veilleux2020}. After star-formation ignites, the most massive stars emit high-energy photons, drive strong stellar winds, and later explode as energetic supernovae \citep[e.g.,][]{Efstathiou2000, Hopkins2012, Hu2019}. These processes inject energy and momentum into the interstellar medium (ISM), heating and accelerating gas in the form of galactic outflows \citep[e.g.,][]{Muratov2015, Nelson2019}. {\ASL Both at low and high-redshifts,} evidence of star-formation driven winds have been found in the {\arf cool, absorbing gas ($n_e \simeq 10^{-2} - 1\ \mathrm{cm^{-3}}$) traced by the low-ionization UV transitions} \citep[e.g.,][]{Weiner2009, Steidel2010, Erb2012, Martin2012, Rubin2014, Roberts-Borsani2020, Prusinski2021, Calabro2022}, as well as in the {\arf warmer ionized gas ($n_e \sim 10 - 10^3\ \mathrm{cm^{-3}}$) probed by nebular emission in the rest-optical} \citep[e.g.,][]{Amorin2012b, Arribas2014, Freeman2019, Gallagher2019, Swinbank2019, Concas2022, ReichardtChu2022, Marasco2023, Llerena2023, Weldon2024}. Outflows are ubiquitous in local, compact starburst galaxies with intense star-formation activity \citep[e.g.,][]{Heckman2002, ReichardtChu2025}, showing velocities of hundreds of ${\rm km ~s^{-1}}$ \citep[e.g.,][]{Martin2005, Chisholm2015, Xu2022-outflows}, and mass outflow rates comparable to their SFR \citep[e.g.,][]{Rupke2005, Heckman2015}. The incidence of outflows plays a major role in galaxy populations near the Cosmic Noon \citep{Newman2012, Genzel2014, Davies2019, FS2019}, partially due to the increase in the cosmic star-formation rate density  \citep[see][for a review]{FS2020}. 

Galactic outflows have the ability to enrich the Circumgalactic Medium (CGM) with metals \citep[e.g.,][]{Muratov2017, Chisholm2018, HamelBravo2024}. In a virialized system, the escape velocity is inherently linked to the dynamical mass \citep[e.g.,][]{Heckman2000, Arribas2014} of the galaxy. Therefore, the shallow potential wells in low-mass galaxies enable their outflows to more easily escape \citep[e.g.,][]{Carniani2024, Xu2025}. {\ASL The dynamical mass also determines the rotational velocity of the gas particles within the galaxy.} Observations of star-forming galaxies at $0 \leq z \leq 3$ \citep[e.g., see][and references therein]{Ubler2019} suggest an exponential increase in gas turbulence with look-back time, which may eventually prevent the formation of rotationally-supported structures in high-$z$ systems \citep[e.g.,][]{Pillepich2019}. {\ASL The dynamical mass, together with galactic feedback \citep[e.g.,][]{Krumholz2018}, fundamentally impacts the morphology and mass assembly history of galaxies \citep[e.g.,][]{Cappellari2006, Belli2014}.} However, empirical constraints on the gas kinematics and the dynamical mass of galaxies beyond Cosmic Noon are limited to only a few studies \citep[e.g., see][]{Parlanti2023, deGraaff2024}.

{\ASL Prior to the commissioning of {\it JWST},} our knowledge about ionized gas flows in galaxies was vastly restricted to $z < 4$, as either the UV spectral features become too faint beyond these distances, or the emission lines in the rest-optical wavelength range become inaccessible from the ground. Luckily, and thanks to the sensitivity and spectral coverage of the NIRSpec instrument \citep[e.g.,][]{Jacobsen2022} on board of the James Webb Space Telescope \citep[e.g.,][]{JWST2023, Rigby2023}, we are now able to study the properties of gas flows and their host galaxies in the first few Gyr of cosmic history \citep[e.g.,][]{Tang2023, Carniani2024, Roy2024, Zhang2024, Xu2025}. {\ASL Studying both the dynamical properties of galaxies and the demographics of outflows at high-redshift is crucial to test the validity of the feedback models \citep[e.g.,][]{Pandya2021} and the formation of structures at early stages of galaxy evolution \citep[e.g.,][]{Kohandel2024}.} In this work, we make use of high-resolution NIRSpec spectroscopy to study the kinematics of the ionized gas, the dynamical mass, and the outflow properties of a sample typically luminous galaxies beyond Cosmic Noon. 

The paper is organized as follows. In Section \ref{sec:data}, we present the GO1871 observations and data reduction, spectral energy distribution (SED) and morphological fitting. We also describe the line emission and outflow modeling. {\ASL In Section \ref{sec:sample}, we present an overview of the properties of the sample in the context of other measurements at high-$z$.} Section \ref{sec:dyn_mass} reads about the velocity dispersion of the ionized gas and the dynamical mass of galaxies beyond Cosmic Noon. Section \ref{sec:outflows} focuses on the properties of the outflow candidates identified in our high-$z$ galaxy sample, concretely on outflow velocities, mass outflow rates, and mass loading factors. In the same section, we link the properties of the outflow to the dynamical mass, aiming to study the efficiency of the feedback and subsequent enrichment of the CGM. We list our conclusions in Section \ref{sec:conclusions}. 

Every discussion across the manuscript is accompanied by comparisons with other samples in the literature, and sometimes with numerical predictions from cosmological simulations. Throughout we assume a cosmology of $\{H_0, \Omega_M, \Omega_{\Lambda}\} = \{70~{\rm km~s^{-1}~Mpc^{-1}}, 0.3, 0.7\}$, speed of light $c = 2.99792 ~10^5 ~{\rm km~s^{-1}}$ and the AB magnitude system \citep{OkeGun1983}. 

\section{Data and methods}\label{sec:data}

\begin{figure*}
    \includegraphics[height=3cm, page=1]{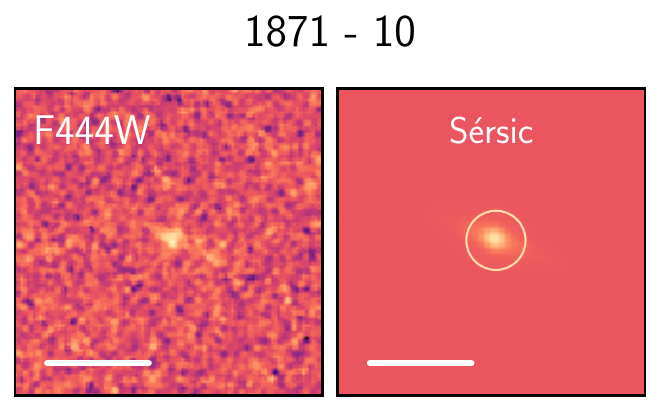}
    \includegraphics[height=3cm, page=1]{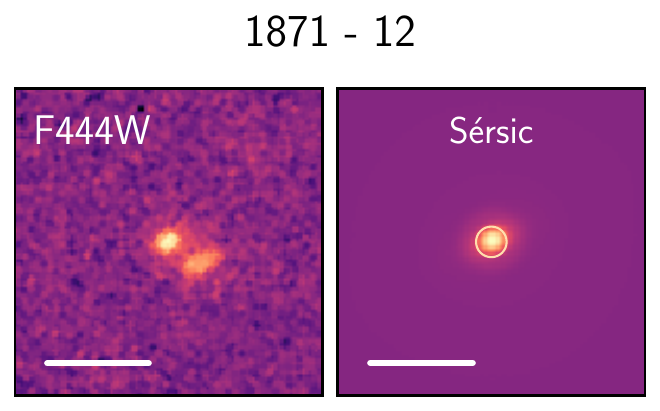}
    \includegraphics[height=3cm, page=1]{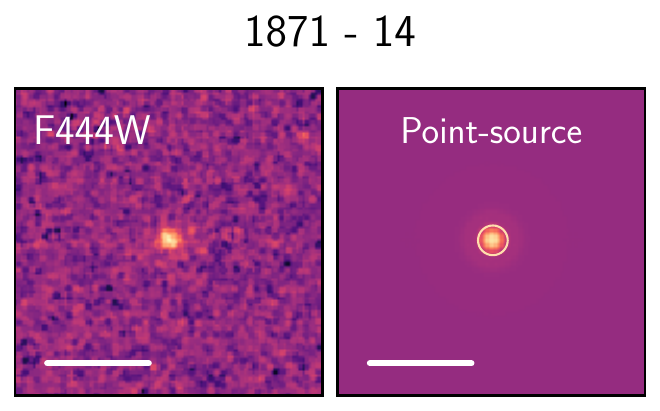}
    \includegraphics[height=3cm, page=1]{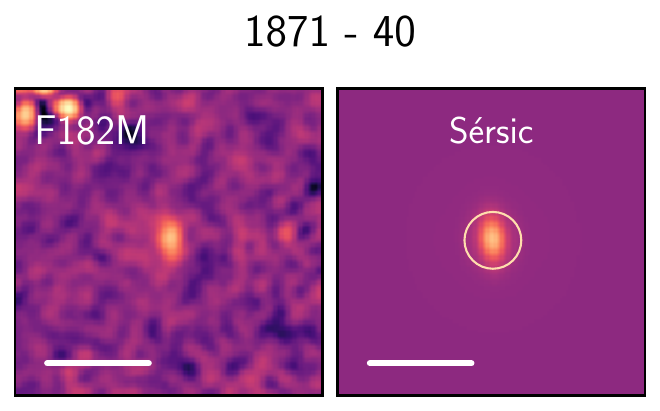}
    \includegraphics[height=3cm, page=1]{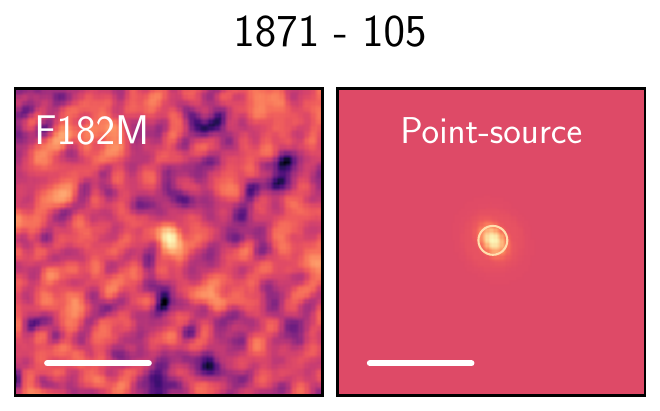}
\caption{{\bf Examples of NIRCam data and morphological modeling for GO1871 gas-flow candidates.} For each galaxy the ID is given, and the \emph{left panels} show $3\times 3$~arcsec$^{2}$ cutouts in the corresponding NIRCam band, with the white bar spanning 1~arcsec length. S\'ersic or point-like best-fit models are shown in the \emph{right panels}, using the \textsc{pysersic} code (see Sect.\ \ref{sub:nircam}). Yellow circles indicate the best-fit effective radius ($r_e$). 
% , or twice the PSF FWHM for unresolved sources.
}
\label{fig:images}
\end{figure*}

\subsection{The GO1871 {\it JWST} program}\label{go1871}
The GO1871 Cycle 1 (PI: Chisholm) program was initially targeted at investigating 20 high redshift star-forming galaxies in the GOODS-North (GOODS-N) field, using {\it JWST} NIRSpec Micro-Shutter \citep{JWST2023, Rigby2023} in both G235H/F170LP and G395H/F290LP gratings. The goal of the project was to study the production and escape of ionizing photons during the Reionization era. With this aim, we used high-resolution gratings to capture velocity-resolved \ion{Mg}{ii} emission profiles at 2800\AA\ rest-frame, which provide important insights into the neutral gas and escape of ionizing radiation from galaxies \citep[$f_{\rm esc}$][]{Henry2018, Chisholm2020, Xu2022-MgII}. {\ASL The main results of this program were presented in \citet{Gazagnes2025}, where we studied the contribution of two luminous galaxies to the ionizing photon budget during the Epoch of Reionization (EoR), by placing indirect constraints on $f_{\rm esc}$ using the \ion{Mg}{ii} line.}

These galaxies were selected from an original catalog of 1,036 sources \citep{Finkelstein2015, Bouwens2015} using deep Hubble Space Telescope (HST) F160W observations to identify $z \geq 5$ targets, with either photometric or spectroscopically confirmed redshifts \citep{Jung2020}. From the parent sample, we kept candidates with $m_{\rm F160W} \leq 28$ AB magnitudes and significant ($3\sigma$) {\it Spitzer} 4.5~$\mu$m detections, ensuring that they are EoR galaxies but bright enough to detect \ion{Mg}{ii} in emission. We used the 100~mas pixel-scale Complete Hubble Archive for Galaxy Evolution \citep[CHArGE; ][]{Fujimoto2022, Kokorev2022}\footnote{\url{https://gbrammer.github.io/projects/charge/}} imaging in the {\it HST} F160W band to define the source locations. Centering the MSA footprint on a bright $z = 7.5$ Ly$\alpha$ emitter in GOODS-N \citep{Finkelstein2013, Hutchison2019, Jung2020}, left us with a final sample of 20 star-forming galaxies successfully allocated inside the micro-shutters. 

\subsection{NIRCam photometry}\label{sub:nircam}
\begin{figure*}
    \includegraphics[height=5.5cm, page=1]{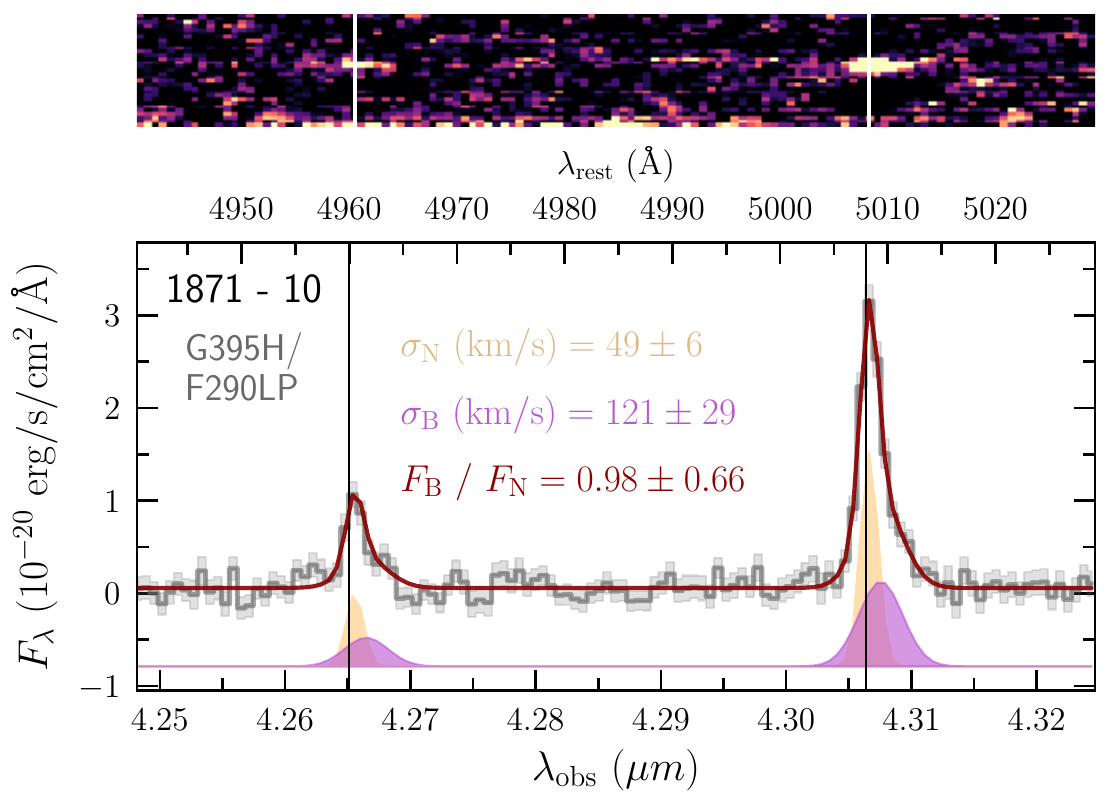}
    \includegraphics[height=5.5cm, page=1]{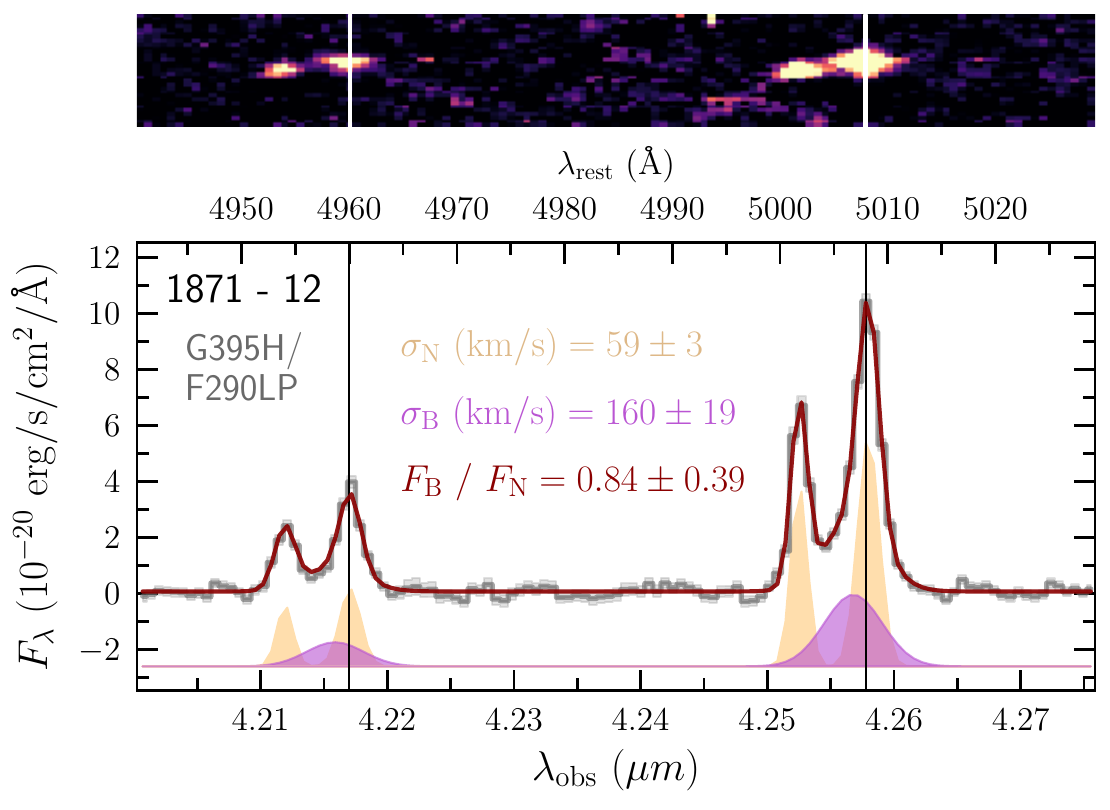}
    \includegraphics[height=5.5cm, page=1]{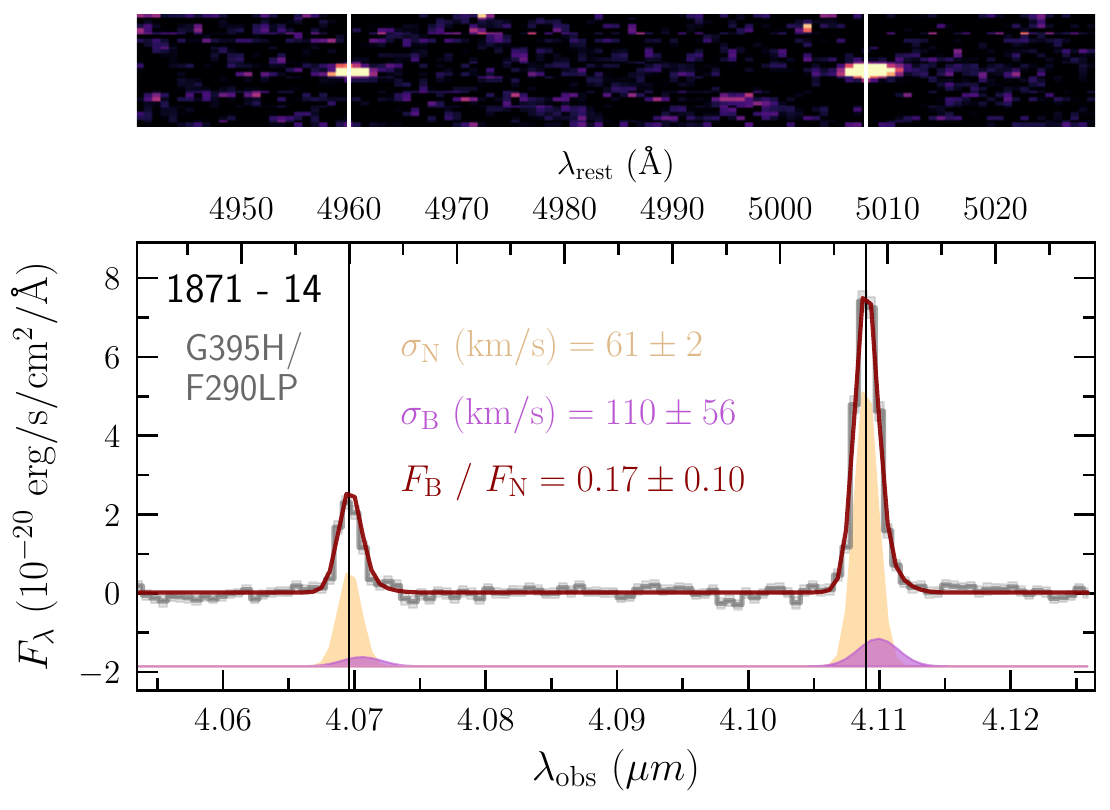}
    \includegraphics[height=5.5cm, page=1]{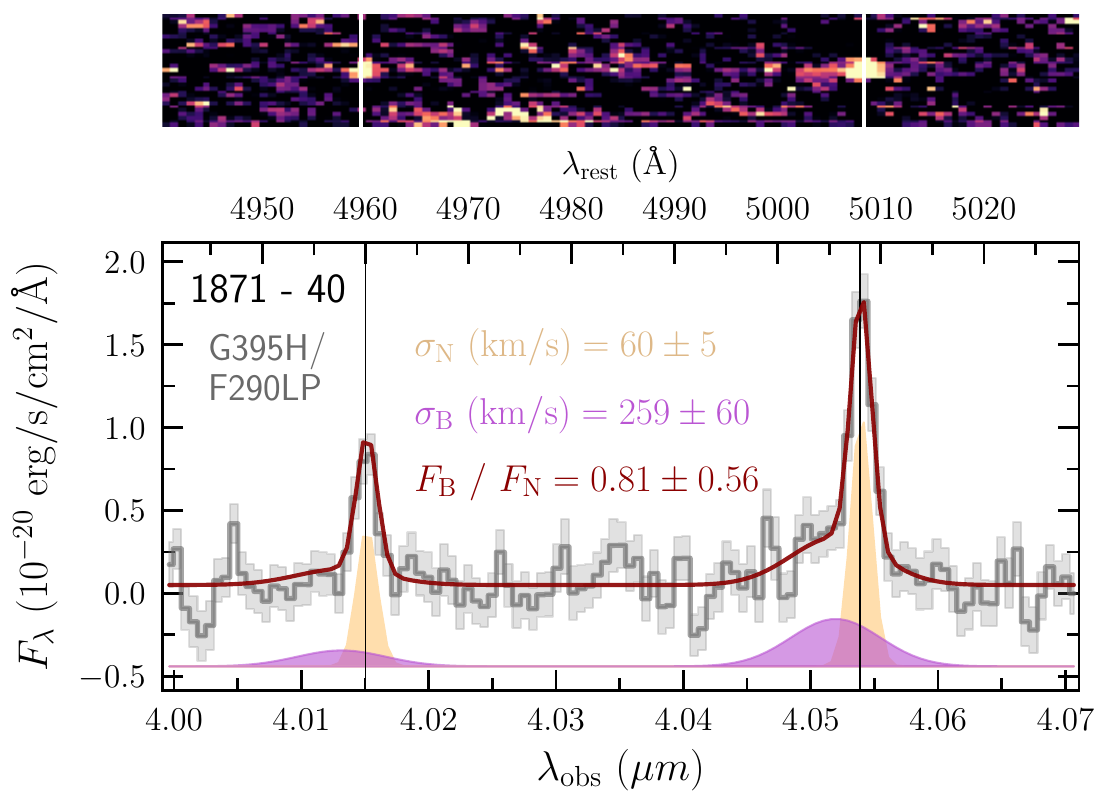}
    \includegraphics[height=5.5cm, page=1]{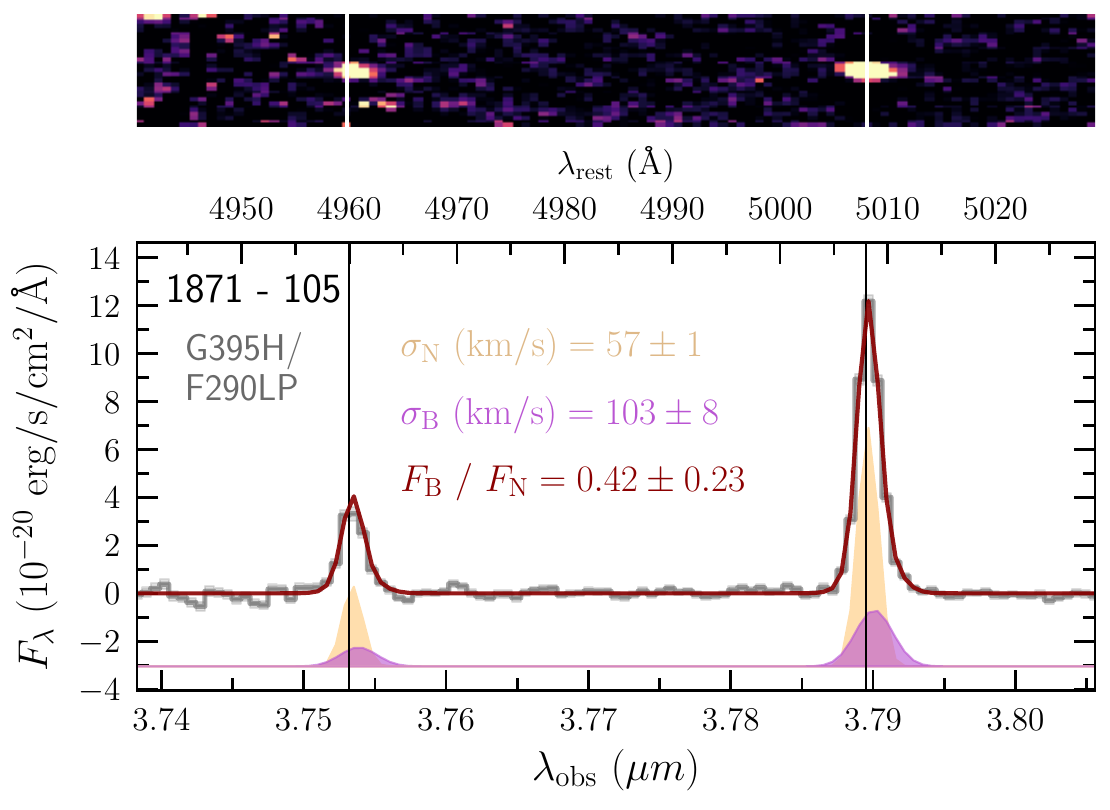}
\caption{{\bf Examples of NIRSpec spectra for GO1871 gas-flow candidates.} Each panel contains the high-resolution G395H~/~F290LP 2D (\emph{top}) and extracted 1D spectra (\emph{bottom}), showing the[\oIII]$\lambda\lambda$4960,5008 doublet (indicated by white and black vertical lines at the top and bottom of each panel,  respectively). The red lines show the best simultaneous fits to both emission lines plus the continuum (see Sect.\ \ref{sub:nirspec}). The narrow, yellow Gaussians reproduce the static component of the ionized gas, and purple Gaussians model the gas-flows over the broad wings of the [\oIII] lines. Best-fit values for the velocity dispersion of each component ($\sigma_{\rm narrow}, \sigma_{\rm broad}$), as well as the broad-to-narrow flux ratios ($F_{\rm B}/F_{\rm N}$) are labeled in the insets. {\ASL Galaxy $1871-12$ has an associated fainter companion at 0.36~pMpc \citep{Gazagnes2025}: two pairs of [\oIII] doublets are clearly resolved in the NIRSpec spectra.}}
\label{fig:spectra}
\end{figure*}

\subsubsection{NIRCam data}
In addition to our dedicated NIRSpec spectroscopy, {\it JWST} NIRCam \citep{nircam} observations of the GOODS-N field were conducted as part of the First Reionization Epoch Spectroscopically Complete Observations \citep[FRESCO; ][]{Oesch23}. These included medium-band observations in the F182M and F210M bands (4,456 and 3,522s, respectively) and shallower F444W broad-band imaging (934s), aiming for a point-source $5\sigma$ detection of 28.2mag in all exposures. The GOODS-N field also benefits from extensive multi-wavelength data, including deep {\it HST} and other ground- and space-based observations \citep[e.g.,][]{Barro2019}.

The FRESCO NIRCam images were processed following the methods in \citet{Endsley2024}, using the {\it JWST} Science Calibration Pipeline\footnote{\url{https://jwst-pipeline.readthedocs.io/en/latest/index.html}} (v1.11.3) {\ASL and the \texttt{jwst\_1106.pmap} reference file.} All NIRCam mosaics were resampled onto the same World Coordinate System at 30 mas pixel$^{-1}$. Images were convolved to match the point-spread function (PSF) of the F444W filter, using empirically derived PSFs from GAIA confirmed stars in the FRESCO mosaics. {\ASL Finally, the photometry was calculated in elliptical Kron apertures after employing a neighbor subtraction algorithm following \citet{Endsley2024}.}

\subsubsection{SED fitting and morphological modeling}
We derived constraints on the SED properties (namely stellar masses, \mstar, and star-formation histories, SFHs) of {\ASL the 20 on-slit GO1871 galaxies} by fitting the {\it HST} plus FRESCO NIRCam photometry with the Bayesian Analysis of Galaxies for Physical Inference and Parameter EStimation code \citep[\textsc{bagpipes}\footnote{\url{https://bagpipes.readthedocs.io/en/latest/}};][]{Carnall2018}. \mstar\ and SFHs were determined by constraining the underlying rest-frame UV to optical continuum, as well as the strength of the prominent [\ion{O}{iii}] and H$\beta$ lines in the redder part of the spectra. {\ASL We accounted for the contribution from the emission lines by including the observed emission line equivalent widths (EW) from the NIRSpec spectra (see \autoref{sub:nirspec}) in the \textsc{bagpipes} fits.}

\textsc{bagpipes} uses the updated \citet{BC03} stellar population synthesis templates with a \citet{Kroupa2001} stellar initial mass function, and includes nebular continuum and emission lines by processing the stellar emission through \textsc{cloudy} v17.00 \citep{Ferland2017}. In fitting the data, we adopted the ``bursty continuity'' prior for the SFHs \citep[following][]{Tacchella2023}, motivated by the large optical equivalent widths of strong emission lines in GO1871 spectra \citep[e.g.,][]{Atek2022}. We allowed for a broad range of stellar masses, metallicities, and ionization parameters, applying log-uniform priors to all three physical properties. We adopt a \citet{Calzetti2000} dust-attenuation law with the V-band optical depth allowed to vary between $A_V = 0.001 - 2$ mag. The redshift was fixed to the NIRSpec spectroscopic value (see below). The resulting SED-derived UV magnitudes, stellar masses, and SFRs of the full sample are listed in Table \ref{tab:data_sample} of the Appendix. 

Finally, we made use of the \textsc{pysersic}\footnote{\url{https://pysersic.readthedocs.io/en/latest/index.html}} software to characterize the morphology of every galaxy \citep{Pasha2023}. We first produced $3\times ~3$~arcsec$^{2}$ cutouts from the PSF-matched F182M, F210M and F444W FRESCO images. We then used \textsc{pysersic} to fit 2D S{\'e}rsic profiles to the light distribution of the F444W band, accounting for PSF convolution and assuming flat priors for the S{\'e}rsic parameters. If a galaxy was not detected at $3\sigma$ in F444W, we used the F182M or F210M cutouts instead (this happens for seven out of 20 galaxies in the sample). \textsc{pysersic} returns the best-fit values for the amplitude, effective radius ($r_e$), S{\'e}rsic index ($n$), ellipticity ($q$) and position angle (PA), using Bayesian Inference. {\ASL In four cases, $r_e$ was smaller than the PSF FWHM (e.g., 0.145~arcsec for F444W). Consequently, we considered these galaxies to be \emph{unresolved}}, and modeled their light profile assuming a point-like morphology instead, by re-scaling the intensity of the PSF to match the galaxy flux. 

Figure \ref{fig:images} shows examples of NIRCam data and the corresponding S{\'e}rsic or point-source best-fit models for a selection of GO1871 galaxies {\ASL (gas-flow candidates in the next Section).} The results from our morphological modeling can be found in Table \ref{tab:data_morphology} of this manuscript. {\ASL It is worth noticing that the $1871-12$ system, first published in \citet{Gazagnes2025}, comprises two galaxies separated by 0.36~pMpc. Hereafter, we will only consider the morphology of the brighter (NE) component, shown at the center of the F444W cutout in the former figure. {\arf Likewise, source $1871-63$ correspond the intermediate-mass black hole (IMBH) candidate by \citet{Chisholm2024}, also part of the GO1871 sample. This galaxy was proposed as a narrow-line AGN via the detection of the high-ionization [\ion{Ne}{v}]$\lambda$3427 line and, as such, is excluded from the current analysis. We note, however, that $1871-63$ does not show significant outflow features in the spectrum}.}

\subsection{NIRSpec spectroscopy}\label{sub:nirspec}
\subsubsection{NIRSpec data}
The NIRSpec observations were split between the G235H and G395H grating configurations. G235H, targeting the rest-frame 2550--3500~\AA\ for galaxies within the EoR ($5.5 < z < 9.5$), focused on the \ion{Mg}{ii} emission lines and received a longer exposure time: 53,044s (14.7 hours) across 36 exposures. The G395H grating, covering 3400–6000 Å at the same redshift, contains brighter optical emission and required less integration time: 9,716s (2.7 hours) over 6 exposures. Both configurations used NRSIRS2 readout mode and the standard three-shutter nod pattern for background subtraction.

For data reduction, we processed the NIRSpec data with the \textsc{msaexp}\footnote{\url{https://github.com/gbrammer/msaexp}} v0.8.4 Python package \citep{msaexp}, {\ASL using the \texttt{jwst\_1235.pmap} reference file.} This applied 1/f noise correction, snowball detection, bias removal using a median filter, and noise re-scaling from empty parts of the exposure. \textsc{msaexp} also performed parts of the Level 2 {\it JWST} calibration pipeline, including manual background subtraction based on 2D slit cutouts. The pipeline used the Space Telescope Science Institute's v1.14.0 standard for Level 2 calibration, with wavelength calibration based on the NIRSpec instrument model \citep{Lutzgendorf2024}. All exposures were co-added without slit-loss correction. {\ASL Four out of 20 sources originally within the MSA micro-shutters show neither continuum nor line detections, and therefore are removed from the final working sample. Our final working sample then constitues of 16 galaxies.} We refer the reader to \citet{Chisholm2024} and \citet{Gazagnes2025} for more details about the NIRSpec data reductions of this program. 

\subsubsection{Emission line measurements}
From \ion{Mg}{II} at 2800\AA\ to the [\ion{S}{II}] doublet at 6730\AA\, a suite of nebular lines are detected in the blue and red portions of the combined NIRSpec GO1871 spectra (G235H and G395H), covering the rest-frame optical wavelengths at the redshift of our observations. Here is a list of the main features, where the name of each ion is followed by the ionization state as well as the rest-frame wavelength (in vacuum): {\ASL [\ion{O}{ii}] $\lambda\lambda$3727,3729, [\ion{Ne}{iii}] $\lambda$3869, H$\delta$ $\lambda$4102, H$\gamma$ $\lambda$4341, H$\beta$ $\lambda$4862, [\ion{O}{iii}] $\lambda\lambda$4960,5008, [\ion{N}{ii}] $\lambda\lambda$6549,6585, H$\alpha$ $\lambda$6564, [\ion{S}{ii}] $\lambda\lambda$6718,6732.} 

We fit individual Gaussian profiles to each line emission  in the observed $f_\lambda$ frame, following:
\begin{equation}
    F_{\lambda} = \mathcal{C} + \dfrac{F_{\rm tot}}{\sigma_{\lambda}\sqrt{2\pi}} ~\exp{\left( -\dfrac{(\lambda - \lambda_0)^2}{2\sigma_{\lambda}^2} \right)},
\end{equation}

\noindent where $\mathcal{C}$ is the local continuum level, assumed to be a positive constant for each transition. The continuum $\mathcal{C}$, amplitude {\ASL $F_{\rm tot}$ (integrated flux of the line),} observed central wavelength $\lambda_0$ (in $\mu m$), and observed velocity dispersion $\sigma_{\lambda}$ ($\mu m$) are free parameters within the fit. During the fit we assumed uniform priors on all free parameters. All of the spectroscopic multiplets are resolved at the resolution of our observations (nominal resolving power of the $H$-gratings is $R \approx 3,000$). {\ASL In this vein, nebular lines coming from the same ionized gas should have comparable velocity dispersions, but the wavelength-dependent resolution of NIRSpec prevents us from tying their $\sigma_{\lambda}$ together without significant forward modeling of the delivered NIRSpec spectral resolution. Instead, only lines within resolved multiplets that are close in wavelength (and of the same ion) are fitted simultaneously by assuming the same velocity dispersion for all constituents (in wavelength space, for example [\oIII]$\lambda$4960 and 5008)}, fixing the relative $\lambda_0$ ratio according to the vacuum wavelengths listed above. {\ASL The contribution from the stellar absorption in the \ion{H}{i} lines is neglected in our analysis, as we expect this to be low given the high EW of the Balmer lines \citep[e.g.,][]{GonzalezDelgado1999}.}

We employ the Python package \textsc{lmfit}\footnote{\url{https://lmfit.github.io/lmfit-py/index.html}} \citep{lmfit} to recover the solution which minimizes the reduced $\chi^2$ value. Errors on the line profile parameters are estimated by perturbing observed $F_{\lambda}$ fluxes using a normal distribution with a mean of zero and equal to the 1$\sigma$ error of the flux density for the standard deviation at every pixel, then re-fitting the lines over $1000$ iterations per spectrum. The best-fit parameters are taken as the median of each distribution, and the uncertainties correspond to the 0.16 and 0.84 quantiles. For non-detections, we report the 1$\sigma$ value as upper limits in the line flux. 

{\ASL In Section \ref{sec:sample} (see also Table \ref{tab:data_sample}), the intensity ratios of the bright optical lines in the GO1871 sample will be compared against other samples in the literature.} {\ASL The merging nature of $1871-12$ is clearly revealed by the presence of double [\oIII] profiles in the NIRSpec spectra of this source (Fig.\ \ref{fig:spectra}). Hereafter, we will only use the line emission of the brighter (NE) galaxy in the system because this is the only component with clear outflow signatures. Finally, we note that the emission line ratios reported in this work are fully compatible with the ones in \citet{Gazagnes2025} for this galaxy component. Similarly, our emission line measurements are compatible with the values presented in \citet{Chisholm2024} for the IMBH candidate.} 

\begin{figure*}
    \includegraphics[width=0.7\textwidth, page=1]{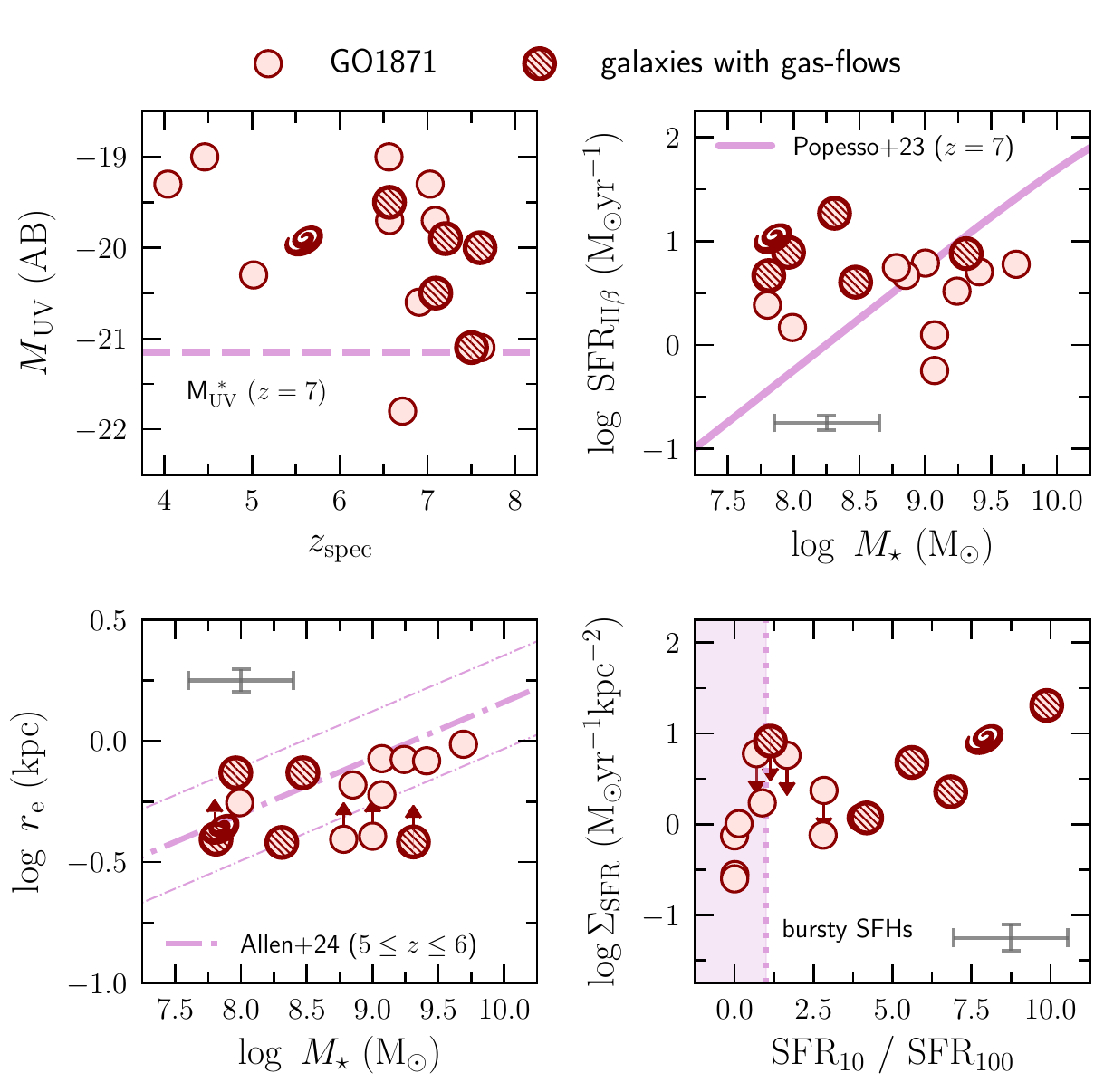}
\caption{{\bf Overview of physical properties for GO1871 galaxies.} Hatched symbols indicate gas-flow candidates, and the IMBH candidate from \citet{Chisholm2024} is displayed with a spiral. \emph{Top-left:} UV magnitude versus redshift. The dashed line shows the characteristic $\muv^*$ for $z=7$ galaxies \citep{Bouwens2021}. \emph{Top-right:} SFR (from H$\beta$) as a function of stellar mass, together with the star-forming main sequence at $z=7$ from \citet{Popesso2023}. \emph{Bottom-left:} rest-optical size-mass relation, as traced by the effective radius ($r_e$) in comparison with the \citet{Allen2025} best-fit relation at $5 \leq z \leq 6$ (dashed-dotted). \emph{Bottom-right:} SFR surface-density (\sigSFR) versus burstiness parameter (SFR$_{10}$/SFR$_{100}$), defined as the ratio between SFRs averaged over 10 and 100~Myr. {\ASL The properties of GO1871 galaxies are representative of typical high-$z$ galaxies in the EoR.}}
\label{fig:sample}
\end{figure*}

\subsubsection{Identifying gas flows}
The imprint of inflows and outflows of ionized gas on the profile of nebular emission lines is in the form of spectral broadening of the emission line wings \citep[e.g.,][]{Heckman1981, Veilleux1991}. Our GO1871 observations are ideal to unveil the presence of inflowing or outflowing gas in the $z \geq 5$ Universe due to the following reasons. First, they target sufficiently bright emission lines, such as [\oIII]5007 or H$\alpha$, the same lines traditionally used to study the kinematics of ionized gas in galaxies and AGNs at lower $z$'s \citep[e.g.,][]{VillarMartin2011, Amorin2012b, Arribas2014, Harrison2014, Gallagher2019, Revalski2021, RuschelDutra2021, ReichardtChu2022, Marasco2023, Flury2023, Amorin2024, ReichardtChu2025}. Second, {\ASL the spectral resolution of G1871 observations ($\simeq 30{\rm ~km~s^{-1}}$) is high enough to separate the gas flow component from the static emission, with typical line widths of $\simeq 60{\rm ~km~s^{-1}}$.}

These broad wing gas-flow features have been traditionally studied through a variety of methods: {\ASL from non-parametric (also called \emph{empirical}) methods \citep[e.g.,][]{Veilleux2005} and detailed numerical models \citep[e.g.,][]{Krumholz2017}, to the inclusion of additional components in the form of Gaussian \citep[e.g.,][]{Pelat1980, Pelat1981}, power-law \citep[e.g.,][]{Komarova2021}, Gauss-Hermite \citep[e.g.,][]{Riffel2010} or even semi-analytic, physically motivated functions to describe the line profile \citep[e.g.,][]{Flury2023}. Due to the limited ${\rm S/N}$ and resolution of our NIRSpec spectra, we search for gas flow signatures by {\ASL fitting the profiles of the bright [\oIII]$\lambda\lambda4960,5007$ doublet} using a simple, 2-Gaussian component approach.} 

In this formalism, an additional Gaussian (name \emph{broad}) is added to the (often brighter) emission coming from the bulk of the static ionized gas (\emph{narrow}), as in:
\begin{equation}
    \begin{split}
    F_{\lambda} = \mathcal{C}~ &+ \dfrac{F^{\rm narrow}_{\rm tot}}{\sigma_{\lambda}\sqrt{2\pi}} ~\exp{\left( -\dfrac{(\lambda - \lambda_0^{\rm narrow})}{2\sigma_{\lambda}^{2, {\rm narrow}}} \right)} ~+ \\
    & + \dfrac{F^{\rm broad}_{\rm tot}}{\sigma_{\lambda}\sqrt{2\pi}} ~\exp{\left( -\dfrac{(\lambda - \lambda_0^{\rm broad})}{2\sigma_{\lambda}^{2, {\rm broad}}} \right)},
\end{split}
\end{equation}

\noindent where $F^{\rm broad}_{\rm tot}$, $\lambda^{\rm broad}_0$ and $\sigma^{\rm broad}_{\lambda}$ correspond to the integrated flux, central wavelength and velocity dispersion of the gas flow, respectively. We force $\sigma^{\rm broad}_{\lambda} \geq 1.5\times \sigma^{\rm narrow}_{\lambda}$, so that the velocity dispersion of the gas-flow component is at least 50 per cent larger than the static one. {\ASL For both narrow and broad components, the velocity dispersion of lines from the same ion that are close in wavelength (i.e., part of the same multiplet) were tied together (e.g,, [\oIII]$\lambda$4560 and 5008).}

The presence of gas flows in each galaxy \citep[see e.g.,][]{Carniani2024} is determined based on the statistical preference of two Gaussians against a single Gaussian fit. {\ASL For a galaxy to be considered as a gas flow candidate \citep[following][]{Xu2025}, we require the \emph{Akaike Information Criterion} (AIC) to be reduced at least by six (${\rm \Delta AIC \leq 6}$), and a signal-to-noise ratio (${\rm S/N}$) for the integrated flux of the broad component to be ${\rm S/N} \geq 3$. AIC has a smaller penalty for the number of free parameters than the \emph{Bayesian Information Criteria} (BIC), and therefore it may include weaker gas-flow signatures. By imposing these criteria, we make sure that our gas-flow candidates are robust, where the resulting $\chi_{\nu}^2$ and BIC both decrease at the same time. This way, we found five gas-flow candidates out of 16 galaxies with [\oIII] observations in GO1871 (details in Sect.\ \ref{sec:outflows}). We do not detect gas flow features in any other transition than [\oIII]. This is most likely be due to ${\rm S/N}$ limitation, as [\oIII]$\lambda5008$ this is brightest transition in our GO1871 spectra}. 

Figure \ref{fig:spectra} shows examples of NIRSpec G395H~/~F290LP spectra for some of the gas-flow candidates in this work, highlighting the \emph{outflow} and \emph{static} Gaussian best-fit components of the [\oIII] lines (purple and yellow, respectively). The broad-to-narrow amplitude ratios, as well as other secondary gas-flow parameters, are collected in Table \ref{tab:data_gasflow} of this paper. 

\section{Sample overview}\label{sec:sample}
The GO1871 working sample is composed of 16 moderately faint ($\muv = -19$ to $-21$ AB) star-forming galaxies at $z = 4 - 7.6$, with a median spectroscopic redshift of $\langle z_{\rm spec} \rangle = 6.8$. Out of these 16 galaxies, 12 (75 per cent of the sample) are EoR galaxies in the redshift range $6 \leq z \leq 8$ and among those, all except one ($1871-912$) are fainter than the characteristic $\muv^{*}=-21.2$ at $z \simeq 7$ \citep[see][]{Bouwens2021}. Figure \ref{fig:sample} depicts a summary of the physical properties of GO1871 galaxies, with the top-left panel showing the UV magnitude versus redshift. The IMBH candidate by \citet{Chisholm2024} and the gas-flow candidates of this work, are highlighted with different symbols (see legend). 

Balmer decrements were used to estimate the amount of (nebular) dust attenuation in GO1871 galaxies. From H$\alpha$ to H$\delta$, we computed every possible Balmer line ratio, and kept those where both lines were detected. The observed Balmer ratios were then compared against theoretical expectations from {\ASL Case B recombination \citep{StoreyHummer1995}, although we acknowledge that deviations from Case B might be expected \citep[e.g.,][]{Scarlata2024, McClymont2025} if the electron densities drastically increase with redshift \citep[e.g.][]{Isobe2023, Reddy2023}}. As a result, all but two galaxies within the GO1871 sample show \emph{negligible} amounts of dust attenuation {\ASL ($\ebv \leq 0.05$, i.e., the measured ratios are consistent with Case B at 1$\sigma$).} For the remaining two galaxies (1871-29, 1871-545) we get $E_{\rm B-V} = 0.049, 0.210$~mag., and their line fluxes were corrected by this amount assuming the \citet{Reddy2015} {\ASL extinction} law. 

SFRs were derived from the H$\beta$ luminosity adopting a \citet{Kroupa2001} IMF, as in \citet{KennicuttEvans2012}. SFRs range ${\rm SFR_{H\beta} / \msun yr^{-1}} = 0.3 - 30$ {(consistent with SFR$_10$ from SED fitting)}, with SED-derived stellar masses (Sect.\ \ref{sub:nircam}) of $\log \mstar/\msun = 7.75-9.75$. The SFR versus \mstar\ diagram (top-right panel of Fig.\ \ref{fig:sample}) revels most of GO1871 galaxies are highly star-forming, where all but five galaxies in the sample fall either well-above or on top of the SF main sequence (MS) at $z=7$ \citep{Speagle2014, Popesso2023}. Notably, most of the galaxies with detected broad components (hatched circles) lie significantly above the SFMS. Concretely, the IMBH and the five gas-flow candidates have low masses ($\log \mstar/\msun \leq 8.5$) and most show positive $\Delta {\rm MS} \geq 0.5$~dex. 

The effective radii of GO1871 galaxies are in the range of $r_e = 0.08-0.18$ arcsec, or $400-960$~pc, with 12 out of 16 galaxies being resolved by the NIRCam imaging. The so-called size-mass relation is presented in the bottom-left panel of Fig.\ \ref{fig:sample}, with the recent \citet{Allen2025} relation at $5 \leq z \leq 6$ overlaid. {\ASL Our galaxies fall within the observational uncertainties of this relation.} In the same manner, star-formation rate surface densities ($\sigSFR {\rm / \msun yr^{-1} kpc^{-2}} = 0.3-20$) were computed by dividing half of the H$\beta$-derived SFRs {\ASL (as only half the flux is contained within $r_e$)} by the area given by a circumference of $r_e$ in radii, as in:
\begin{equation}
    \sigSFR = \frac{\rm SFR_{H\beta}}{2 \pi r_e^2},
\end{equation}

Our custom \textsc{bagpipes} SED fitting allows us to compute SFRs over different time spans for each galaxy. Commonly used time bins are ${\rm SFR_{10}}$ and ${\rm SFR_{100}}$, averaged over the last 10 and 100~Myr, respectively \citep{Endsley2024}. By definition, ${\rm SFR_{10}}$ traces the most recent (instantaneous) SFR events and closely follows the values derived from the H$\beta$ recombination line (${\rm SFR_{H\beta}}$). Likewise, ${\rm SFR_{100}}$ is largely driven by the luminosity of the UV continuum, and it traces a more continuous level of the the SFR over time. As a consequence, the ratio between both can be interpreted as a proxy of recent burstiness in the SFH of galaxies \citep{Atek2022}. The last panel of Fig.\ \ref{fig:sample} displays the positive correlation between the \sigSFR\ and this `burstiness' parameter. Interestingly, the Chisholm et al. IMBH and most of our gas-flow candidates show high \sigSFR\ (highly star-forming and compact) as well as recent bursts of star-formation (${\rm SFR_{10} / SFR_{100}}$ > 1). 

\begin{figure}
    \includegraphics[width=0.95\columnwidth, page=1]{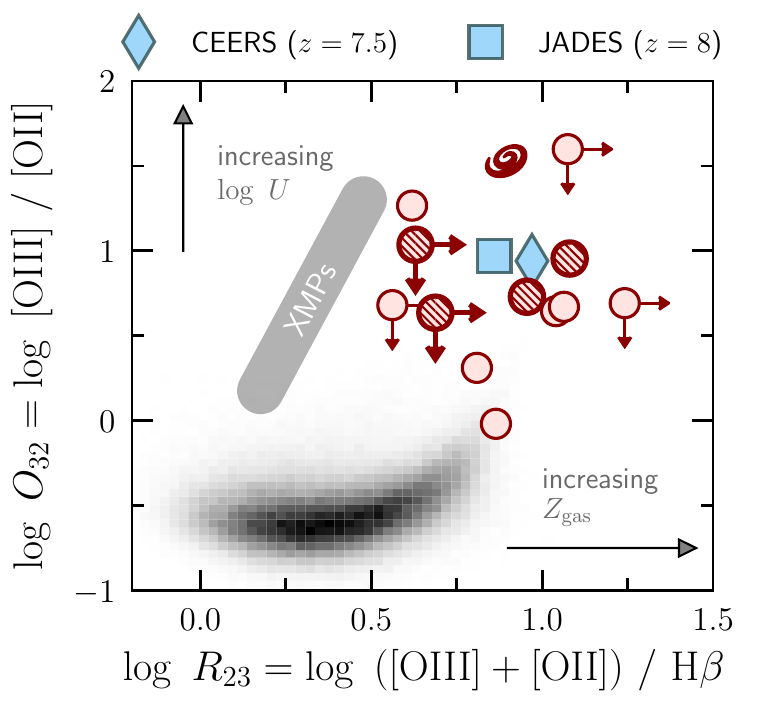}
\caption{{\ASL {\bf $O_{32}$ (a proxy of the ionization state) versus $R_{23}$.} The blue diamond and square show the average for $z = 7-8$ galaxies from the CEERS \citep{Sanders2023} and JADES surveys \citep{Cameron2023}. The grey 2D histogram in the background shows the density of SDSS galaxies at $z \simeq 0$ \citep[e.g.,][]{Brinchmann2008}. The space of parameters occupied by the compilation of extremely metal-poor galaxies (XMPs) from \citet{Izotov2024}, is illustrated with a grey shaded band. With high ionization parameters and low metallicities, high-$z$ star-forming galaxies share the locus in this diagram with local analogs such as Green Pea, Bluberries and EELGs.}}
\label{fig:excitation}
\end{figure}

{\ASL With the advent of {\it JWST}/NIRSpec, we have access for the first time to optical emission-line diagrams at redshifts higher than $z = 4$ \citep[e.g.,][]{Schaerer2022b, Trump2023, Trussler2023}. We now use the (attenuation corrected) $O_{32} = [\oIII]\lambda5007/[\ion{O}{ii}]\lambda\lambda3727,29$ versus $R_{23} = ([\oIII]\lambda5007+[\ion{O}{ii}]\lambda\lambda3727,29)/H\beta$ diagram to investigate the ionization state of the ionized ISM in our sample of $4 \leq z \leq 8$ galaxies (see Table \ref{tab:data_sample}). $O_{32}$ is a direct proxy of the degree of ionization \citep[e.g.,][]{KewleyDopita2002, Kobulnicky2004, NakajimaOuchi2014}, while $R_{23}$ is primarily sensitive to the gas-phase metallicity with a secondary dependency on the ionization parameter \citep[e.g.,][]{Maiolino2008, Curti2017, Curti2020}.}

{\ASL Figure \ref{fig:excitation} shows the $O_{32}-R_{23}$ diagram for GO1871 galaxies. Our sample of high-$z$ galaxies show overall high $O_{32}$ and $R_{23}$ values ($\log O_{32} = 0.5-2$, $\log R_{23} = 0.5-1.5$) compared to the bulk of nearby SDSS galaxies \citep[grey shaded 2D histogram, e.g.,][]{Brinchmann2008}, compatible with the findings of the literature at Cosmic Noon \citep[e.g.,][]{Strom2017}. These high $O_{32}$ and $R_{23}$ are in agreement with other individual {\it JWST} observations at similar redshifts \citep[e.g,][]{Saxena2024, Mascia2023}, with a tendency for high-$z$ galaxies to move towards the high ionization locus in almost \emph{every} emission line ratio diagnostic diagram \citep{Shapley2023a, Backhaus2024}. As a comparison example, in Fig.\ \ref{fig:excitation} we also show stacked measurements of $z \geq 7$ galaxies from the JADES and CEERS surveys \citep[blue symbols, see][]{Cameron2023, Sanders2023}. In addition, GO1871 galaxies share ionization parameters and excitation ratios with nearby Green Peas \citep[e.g.,][]{Cardamone2009, Amorin2012a, Jaskot2013}, Blueberries \citep[e.g,][]{Yang2017}, Extreme Emission Line Galaxies \citep[EELG, e.g.,][]{Atek2011,Tang2019, Onodera2020, Berg2022classy} and strong Lyman Continuum Emitters \citep[e.g.,][]{Izotov2018, Izotov2020, Nakajima2020, Flury2022} at low and/or intermediate redshifts. Extremely metal-poor galaxies (XMPs), conversely, move to the left-hand side part of this diagram \citep[e.g.,][]{Izotov2021, Izotov2024}, as the metal lines become weaker (see grey shaded band in the plot).} 

{\ASL Taking altogether, these results reveal a coherent picture by which the typical high-$z$ galaxy seems to host a highly ionized and moderately metal-poor ISM with low dust attenuation \citep[cf.,][]{Nakajima2023}. As widely studied in the literature, these conditions are might be driven by the increased hardness of the ionizing spectra in low-metallicity stellar populations \citep[e.g.,][]{Nakajima2016, Steidel2016, Onodera2016, Sanders2016, Strom2017, Strom2018, Sanders2021, Runco2021, Papovich2022}. Last but not least, it is worth noticing the extreme $O_{32}$ observed in the IMBH candidate of \citet{Chisholm2024}. Reproducing the complete nebular structure of this galaxy, including transitions such as [\ion{Ne}{v}] and \ion{He}{ii}, required invoking both massive stars and accretion on to a black hole. For the rest of the GO1871 galaxies we do not observe evidence of AGN largely contributing to the observed properties of the galaxies, neither through broad line regions (BLRs) nor high-ionization lines. These observations suggest that stellar feedback may dominate the baryon cycle of GO1871 galaxies. We stress, however, that the presence of low-mass black holes, leading to a possibly subdominant contribution of AGN feedback, cannot be empirically ruled out.} 

\section{Dynamical masses in high-redshift galaxies}\label{sec:dyn_mass}
The atmospheric sky transmission prevents ground-based telescopes to observe rest-frame optical emission lines beyond $z = 4$. Luckily, the NIRSpec instrument on board of {\it JWST} \citep{Jacobsen2022} offers both the spectral coverage and instrumental resolution to resolve the ionized gas kinematics of star-forming galaxies beyond this redshift. 

\subsection{Velocity dispersion of the ionized gas}
Our Gaussian fits to the GO1871 spectra reveal velocity dispersions of the order of $\sigma{\rm ~(km ~s^{-1})} \simeq 50-100$ for [\oIII]$\lambda$5008, {\ASL aligning, for instance, with previously reported values for the narrow component in nearby starburst galaxies \citep[e.g.,][]{Chavez2014, Amorin2024}.} We use the publicly available \textsc{msafit}\footnote{\url{https://github.com/annadeg/jwst-msafit}} python package \citep{deGraaff2024} to estimate the NIRSpec spectral resolution for every galaxy at the wavelength of the [\oIII] line. \textsc{msafit} is a forward modeling software that accounts for complexities such as the PSF, source size and location with shutters, shutter geometry, bar shadows and pixelation of the NIRSpec data. It uses the source morphology ($r_e, n, q, {\rm PA}$, see Sect.\ \ref{sub:nircam}), the location of the source in the MSA array and its positioning within the slit to compute the \emph{Line Spread Function} (LSF) of the instrument. 

\begin{figure}
    \includegraphics[width=0.95\columnwidth, page=1]{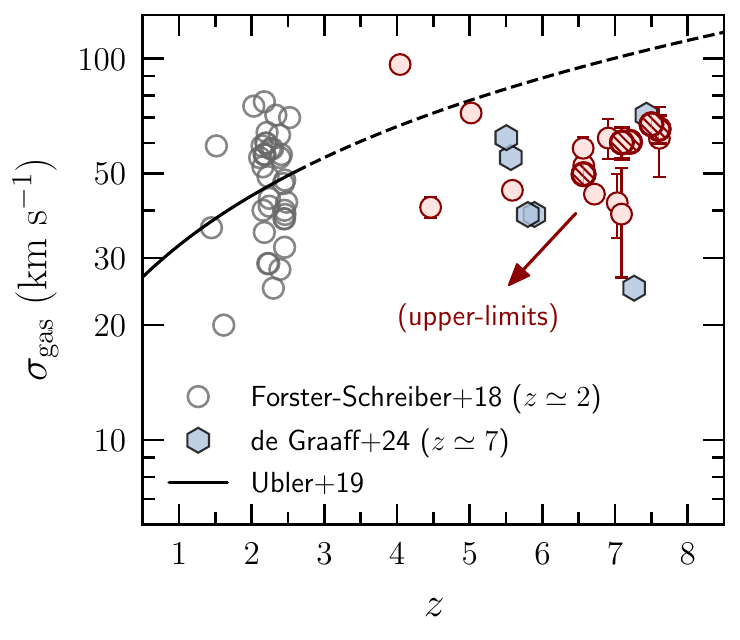}
\caption{{\bf Evolution of the ionized gas velocity dispersion ($\sigma_{\rm gas}$) with redshift.} Open circles show the measurements from \citet{FS2018} at $z \simeq 2$, and the solid (dashed) line indicates the linear fit (extrapolation) by \citet{Ubler2019}. Other NIRSpec studies {\ASL (using [\oIII] and H$\alpha$)} at similar redshifts to this work are shown with hexagons \citep{deGraaff2024}. Ionized gas in high-$z$ SFGs have comparable (or lower) velocity dispersions than Cosmic Noon galaxies.}
\label{fig:sigma_z}
\end{figure}

We estimate $\sigma_{\rm res}{\rm ~(km ~s^{-1})} = 29-38$ for the velocity dispersion introduced by NIRSpec at observed-frame wavelengths of 5008\AA. Then, we correct for the intrinsic velocity dispersion of the ionized gas ($\sigma_{\rm gas}$), accounting for instrumental broadening, as, 
\begin{equation}
    \sigma_{\rm gas} = \sqrt{\sigma^2 - \sigma^2_{\rm res}},
\end{equation}

\noindent resulting in $\sigma_{\rm gas}{\rm ~(km~s^{-1})} = 38-96$ (Table \ref{tab:data_morphology}). GO1871 galaxies seem to have spectroscopically resolved [\oIII] profiles, in which the observed $\sigma$ values are broader than the instrument LSF for \emph{all} systems. Figure \ref{fig:sigma_z} shows the ionized-gas velocity dispersions of our sample as a function of redshift. {\arf It is important to note that $\sigma_{\rm gas}$ not only encompass the thermal and turbulent motions of the gas, but also includes potential rotational components. Accounting for rotational motions will require the proper kinematic modeling of each galaxy \citep[e.g.,][]{deGraaff2024, Danhaive2025}, which is beyond the scope of this paper. Instead, our values for $\sigma_{\rm gas}$ are reported as upper limits to the intrinsic velocity dispersion of the ionized gas (indicated by downward arrows in Fig. \ref{fig:sigma_z} ).} We now compare with the near-IR SINS/zC-SINF sample of \citet{FS2018} at $z \simeq 2$ and higher stellar masses than this work, and with the NIRSpec results presented in \citet{deGraaff2024}, at comparable redshifts and stellar masses. 

Ground-based spectroscopic surveys at $z = 0 - 4$ targeting the kinematics of the ionized gas (SAMI: \cite{Varidel2020}; MAGPI: \cite{Mai2024}; DEEP2: \cite{Kassin2012}; KROSS: \cite{Johnson2018}; SINS/zC-SINF: \cite{FS2018}; KMOS$^{\rm 3D}$: \cite{Wisnioski2015, Ubler2019}; MOSDEF: \cite{Price2020}; KDS: \cite{Turner2017}), have demonstrated that the velocity dispersion of star-forming galaxies increases with redshift. To illustrate this behavior, in Fig.\ \ref{fig:sigma_z} we include the analytic fit to the KMOS$^{\rm 3D}$ data ($\log \mstar/\msun = 9-11$) at $z \simeq 1 - 3$ \citep[solid line, from][]{Ubler2019}. Extrapolating this fit to $z = 7$ (dashed line) suggests that $\sigma_{\rm gas}$ was even higher in high-$z$ systems, {\arf as recently investigated by \citet{Danhaive2025} using NIRSpec observations of a sample of galaxies with slightly higher (average) stellar mass than this work} \citep[see also the first systematic measurements at $z \geq 4$ with ALMA, e.g.,][]{Fraternali2021, Rizzo2021, Parlanti2023, Rizzo2023, Rizzo2024}. Consistently, prevailing galaxy formation theories predict more turbulent galaxies in the early Universe as an outcome of {\ASL increase star-formation}, merger activity, gravitational instabilities, the accretion of gas from the cosmic web, and stellar feedback \citep[e.g.,][]{Krumholz2018, Pillepich2019, Ejdetjarn2022, Kohandel2024}. 

The previously mentioned trend of increasing $\sigma_{\rm gas}$ with $z$ does not extend to the high-$z$ of our galaxies. Both \citet{deGraaff2024} and our GO1871 measurements report average values of the velocity dispersion comparable to Cosmic Noon galaxies. Given that the typical stellar mass of the $z = 7$ samples is 1--2~dex lower than the literature data at lower redshifts, one would expect the $z = 7$  galaxies to have a factor of $\sim$2 lower $\sigma_{\rm gas}$ than we observe. However, the underlying evolution of $\sigma_{\rm gas}$ with $z$ at fixed \mstar\ compensates for this effect \citep[e.g.,][]{Pillepich2019}, with $\sigma_{\rm gas}$ increasing by the same factor due to the above mentioned phenomena. The complementary behavior can be seen at lower redshifts. For instance, at fixed $\log \mstar/\msun = 8-9$ (similar to the \mstar\ of our sample), \citet{Varidel2020} reported average $\sigma_{\rm gas}$ as low as ${\rm \simeq 20 km~s^{-1}}$ for $z=0$ galaxies, while \citet{Maseda2013} measured a mean of ${\rm \simeq 50km~s^{-1}}$ over a sample of $z=2$ dwarfs. In conclusion, our observations suggest that the velocity dispersion of ionized gas increases at fixed \mstar\ with increasing redshift.

% These low $\sigma_{\rm gas}$ values respect to cosmological predictions seem to be in line with recent findings of rotationally-supported and disc-like structures with {\it JWST} \citep[][]{Ferreira2022, Ferreira2023, Conselice2024, Tsukui2025, Xu2024-disk} and ALMA \citep{Rowland2024}, out to much higher redshifts than previously expected \citep[see][for a simulation-based work]{Kohandel2024}. 
% {\ASL According to models \citep[e.g.,][]{Ejdetjarn2022, Rizzo2024}, the stellar mass and SFR are the primary drivers of the ionized-gas velocity dispersion at fixed cosmic time, and observations have established dependencies of increasing $\sigma_{\rm gas}$ with both increasing \mstar\ and SFR \citep[e.g.,][]{Lehnert2009, Green2010, Yu2019}. To illustrate this, in Figure \ref{fig:sigma_mstar} we have plotted the estimated $\sigma_{\rm gas}$ as a function of the galaxy stellar mass, for both high-$z$ \citep[][this work]{deGraaff2024} and the Cosmic Noon samples. Although this diagram do not show any apparent correlation, ... }

\begin{figure*}
    \includegraphics[width=0.7\textwidth, page=1]{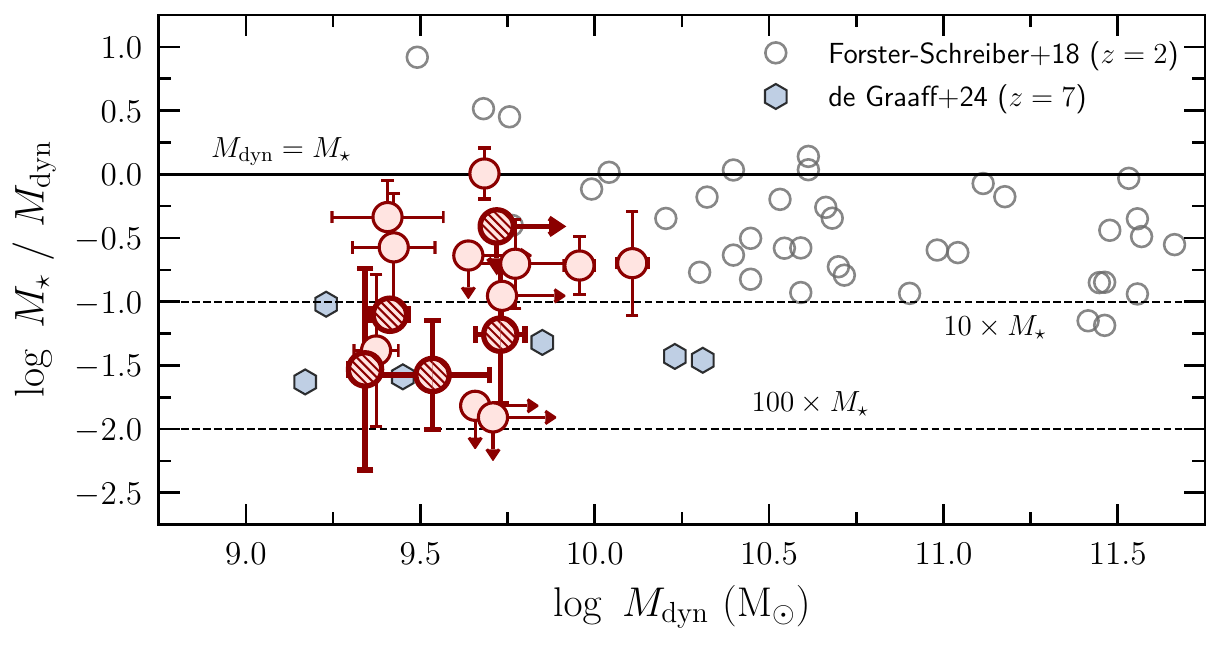}
    \includegraphics[width=0.7\textwidth, page=1]{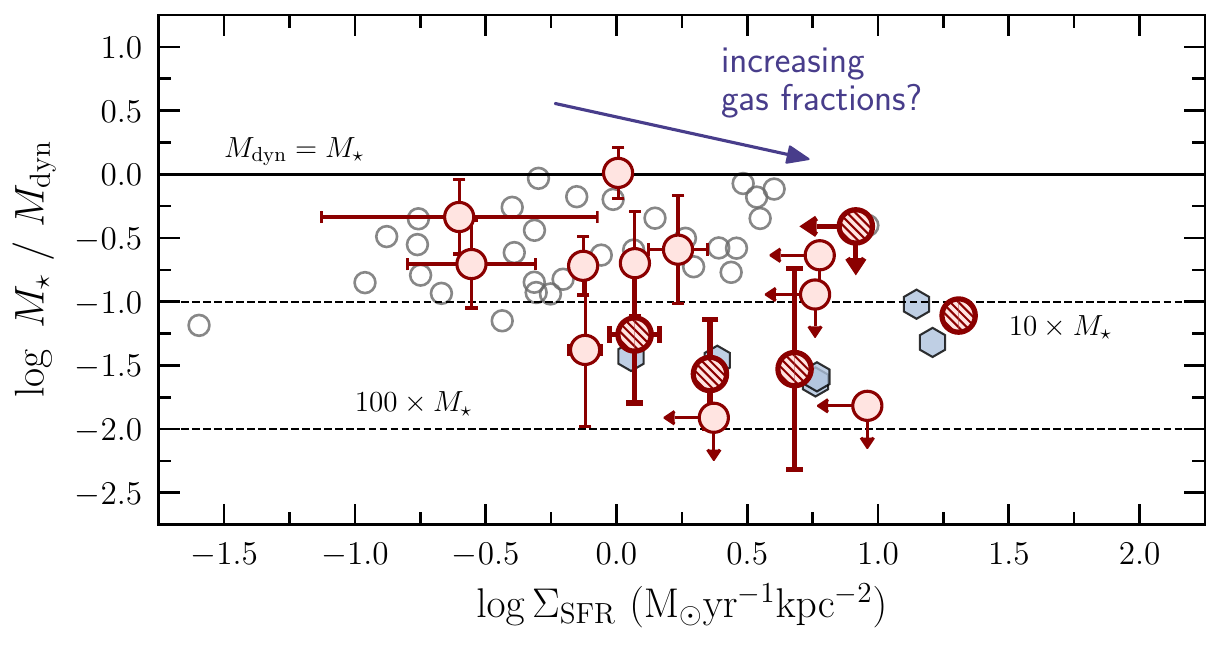}
\caption{{\bf Stellar mass (\mstar) to dynamical mass (\mdyn) ratio as a function of dynamical mass (\emph{top}) and SFR surface density, \sigSFR (\emph{bottom}).} The solid line follows the $\mdyn = \mstar$ relation, while dashed lines show $\times10\mstar$ and $\times 100\mstar$. On average, low-mass, high-$z$ galaxies with moderate SFRs exhibit lower $\mstar/\mdyn$ ratios and lower dynamical masses than more massive and highly star-forming systems at Cosmic Noon (top figure). This behavior can be attributed to higher gas-mass fractions in high-$z$ systems, as suggested by the tentative correlation found between $\mstar/\mdyn$ and \sigSFR\ (bottom figure). Symbols are the same as in Fig.\ \ref{fig:sigma_z}.}
\label{fig:mstar_mdyn}
\end{figure*}

\subsection{Dynamical and baryonic mass budgets}
Taking advantage of the resolved [\oIII] lines in GO1871 galaxies, we now investigate the relation between the dynamical mass (\mdyn) and stellar mass (\mstar). In virialized systems, the mean velocity dispersion of the gas particles is a representation of the average thermal (kinetic) energy, and is therefore linked to the gravitational potential wells responsible for imprinting rotational motions into the gas particles. The mass-equivalent to this gravitational potential is called the \emph{dynamical mass}, and it reflects the addition of the dark and baryonic mass {\ASL enclosed within a certain radius.} 

Following \citet{Ubler2023} and \citet{Maiolino2024}, the dynamical mass (\mdyn) can be defined as, 
\begin{equation}
    \mdyn = K(n) ~K(q) \times \dfrac{\sigma_{\rm gas}^2 r_e}{G},
\end{equation}
\noindent where $K(n) = 8.87 - 0.831n + 0.0241n^2$ from \citet{Cappellari2006}, $K(q) = [0.87 + 0.38~e^{-3.71(1-q)}]^2$ from \citet{vanderWel2022}, and  $n, q, r_e$ are the S{\'e}rsic index, axis ratio and effective radius from our morphological fits to the NIRCam data (see results in Table \ref{tab:data_morphology}). Among the uncertainties on the several quantities involved in the determination of \mdyn, the $r_e$ contributes the most to the final error budget. Four out of 16 galaxies are unresolved in the NIRCam mosaics. For those, we report upper-limits in \mdyn\ by replacing the effective radius with the FWHM of the PSF in the F444W band. {\arf As a persistent caveat, the effect of galaxy inclination is not accounted in the previous equation, and can lead up to 0.3~dex. systematic uncertainties \citep[e.g.,][]{deGraaff2024}.}

The resulting $\mstar/\mdyn$ ratio is plotted as a function of \mdyn\ in the upper panel of Figure \ref{fig:mstar_mdyn}. The dynamical mass ($\log \mdyn/\msun \geq 9.25-10.25$) is larger than the galaxy stellar mass for both GO1871 and de Graaff et al. samples. Conversely, {\arf Cosmic Noon \citep{FS2018} and other high-$z$ galaxies \citep{Danhaive2025} of higher stellar mass,} show larger $\mstar/\mdyn$ ratios and $\log \mdyn/\msun \geq 10$. {\ASL Because the contribution of dark matter to the mass budget is believed to be negligible at these small spatial scales {\arf \citep[$\leq 1~$kpc see e.g.,][]{CourteauDutton2015, Genzel2017}}, the discrepancy between the stellar and dynamical mass is usually attributed to the contribution of the total (ionized+neutral) gas mass to the dynamical mass \citep[$M_{\rm gas}$, e.g.,][]{Erb2006, Chavez2014, Price2016, Wuyts2016}, which has also been postulated as the driver of the black-hole to stellar mass relation at high-$z$ \citep{Maiolino2024}.} This would mean, consequently, that low-mass galaxies at high redshift have higher gas fractions than their massive counterparts, a fact that has been well-established in the $z \geq 1$ literature \citep[][]{Scoville2017, Liu2019, Dessauges2020}. It is worth noticing \citep[see also][]{deGraaff2024}, however, that the magnitude of the discrepancy --an order of magnitude on average, but up to $\simeq 2$~dex in some cases-- is significantly larger than previous lower redshifts studies at similar stellar masses \citep[e.g.,][]{Maseda2013}. {\arf To explain this discrepancy, some authors have suggested that dark matter could dominate the mass-budget in the core of low-mass galaxies \citep[e.g.,][]{Bouche2022}, a result recently supported by some cosmological simulations \citep{deGraaff2024_simulations}. More observational and modeling work is necessary to confirmed or rule out this scenario.}

To test the (more commonly invoked) higher-gas-fraction hypothesis, we look at the gas masses implied by the SFRs of our high-$z$ samples. First, we investigate the correlation between the $\mstar/\mdyn$ and $\sigSFR$ as a proxy of the gas-mass surface density in galaxies \citep[$\Sigma_{\rm gas}$,][]{Kennicutt1998}. {\ASL We found a tentative anti-correlation between $\mstar/\mdyn$ and \sigSFR\ where galaxies with higher \sigSFR\ show lower $\mstar/\mdyn$ ratios (lower panel of the Fig.\ \ref{fig:mstar_mdyn}). A similar correlation between $\mstar/\mdyn$ and $\sigSFR$ was found by \citet{Wuyts2016}, in a sample of massive galaxies at $0.6 \leq z \leq 2.6$. While the SFRs of high-$z$ samples are lower than Cosmic Noon massive galaxies, their star formation is more concentrated (smaller $r_e$), henceforth increasing the \sigSFR. In addition to their low $\mstar/\mdyn$ ratios, this imply higher gas mass fractions in these high redshift systems, in agreement with predictions from cosmological simulations too \citep{Pillepich2019}.}

\emph{Are the implied gas masses enough to reconcile baryonic and dynamical masses?} In order to address this question, {\ASL we use the combined ${\rm \sigSFR-to-\Sigma_{\rm gas}}$ relation (including normal and starburst galaxies) in \citet{Kennicutt2021}} to give a rough estimate of the total gas mass surface density \footnote{ The following gas masses should be taken with caution, as the \citet{Kennicutt2021} relation has been calibrated in the local Universe and assumes a constant star-formation efficiency, which may not apply to high-$z$ galaxies \citep[see][]{Tacconi2020}}. Ultimately, GO1871 galaxies show $\sigSFR / \msun {\rm yr^{-1} kpc^{-2}} = 0.3-20$, which translate into $\Sigma_{\rm gas} / \msun {\rm yr^{-1} kpc^{-2}} = 140-3300$. Accounting for galaxy sizes in the morphologically resolved sources, $r_e = 0.08-0.18$~arcsec (or $400-960$~pc), we obtain gas masses of $\log M_{\rm gas} \simeq 8.2-9.5~\msun$, implying gas-to-stellar mass fractions of $f_{\rm gas} \simeq 0.65$, on average. These large $f_{\rm gas}$ are consistent with the large measured \sigSFR\ values. 

The potential total baryon contribution including the gas phase, $M_{\rm baryon} = \mstar + M_{\rm gas}$, only reconciles baryonic and dynamical masses in 2 out of the 11 (resolved) GO1871 galaxies. For the remaining cases, \mdyn\ is still larger than the total baryonic mass by an inverse factor of $M_{\rm baryon}/\mdyn \simeq 0.2-0.7$. A further increase of the gas masses {\ASL at fixed \sigSFR\ --i.e., an offset in the star-formation law relation--} would require decreasing the star formation efficiency (SFE) in these high-$z$ systems. This scenario, nevertheless, seems unlikely for two reasons. First, there is already an established scaling relation in the local Universe that predicts an increase (or flattening, but never a decrease) of the SFE with \sigSFR\ \citep{Leroy2008}. In addition, such a significant decrease in SFE would go against the claim of a boost in SFE at high-$z$ to explain the over abundance of UV-bright galaxies observed by {\it JWST} \citep[e.g.,][]{Li2024, Dekel2023, Wang2023}. {\ASL Even under the assumption that these galaxies may host an over-massive BH, similar to the recently discovered Little Red Dots \citep[][]{Furtak2023, Kokorev2023, Larson2023, Ubler2023, Lambrides2024} and other broad and narrow-line AGNs at high-$z$ \citep[e.g.,][]{Harikane2023, Chisholm2024, Maiolino2024}, the maximum contribution of the BH to the baryonic mass would not exceed 0.1~dex, insufficient to account for the mass discrepancy. In any case, the absence of broad line regions in the spectra of GO1871 galaxies, and the lack of high-ionization lines (except for the IMBH candidate) makes this scenario unlikely.}

All this said, there is still the possibility of \mdyn\ being overestimated. As widely discussed in \citet{deGraaff2024}, the underlying assumptions in the \mdyn\ equation can only yield an over-estimation of the dynamical mass by $0.3-0.6$~dex, at most. Likewise, the over-estimation of \mdyn\ due to the dominance of outflow components or turbulence due to stellar feedback in the line profiles is unlikely and negligible, respectively \citep[see also discussion in][]{Ubler2019}. Neither we can discard systematic underestimations of stellar mass, due to (1) SFH assumptions \citep[e.g.,][]{Maraston2010, Whitler2023} and (2) the lack of ability of the NIRCam photometry to capture the light from an older (redder) stellar population. This older stellar population may dominate the mass budget but the younger stellar population is out-shining the older stellar population  \citep[e.g.,][]{Papovich2001, Narayanan2024}. Unfortunately, current data do not allow us to draw further conclusions in this direction. 

\section{Gas flows beyond Cosmic Noon}\label{sec:outflows}
Before the launch of {\it JWST}, the properties of gaseous winds beyond Cosmic Noon galaxies remained uncharted territory, relegated to a few number of studies from the ground. These studies targeted the ionized gas traced by rest-UV transitions \citep[e.g.,][]{Sugahara2019}, or the sub-mm regime with ALMA, probing both the cold molecular \citep[e.g.,][]{Jones2019, Spilker2020} and warm ionized phases \citep[e.g.,][]{Gallerani2018, Ginolfi2020, Akins2022}. By exploiting the capabilities of NIRSpec, we now study the properties of warm ionized gas-flows in our sample of $z = 4-8$ galaxies. 

The presence of ionized gas-flows in GO1871 spectra is based on the statistical preference of a two-component versus a single Gaussian fit to the [\oIII]$\lambda$5008 line profile. Following the methods outlined in Sect.\ \ref{sec:data}, we detect five gas-flow candidates out of 16 galaxies with [\oIII] observations. This translates into a 30 per cent detection rate. Although this is in agreement with other NIRSpec studies at similar redshifts \citep[e.g.,][]{Carniani2024}, {\ASL it is still slightly higher than \citet[][who use lower resolution gratings and therefore might be incomplete]{Xu2025}.} GO1871 gas-flow candidates have $\log \mstar/\msun = 7.75-9.25$ and ${\rm SFR_{10} /\msun yr^{-1}} \simeq 10$ at redshifts of $z=6.5-7.5$ (i.e, deep into the EoR). The gas-flow candidates have high \sigSFR\ values of $\sigSFR / \msun{\rm yr^{-1} kpc^{-2}} \simeq 1$, and show evidence of recent bursts in the SFH, as suggested by their high burstiness parameters (${\rm SFR_{10}/SFR_{100} \geq 1}$). {\ASL The gas-flow galaxies have stellar masses below the median of the full GO1871 sample, while their SFR, \sigSFR, and burstiness parameter are high compared to the GO1871 galaxies that do not show gas-flow signatures. {\arf As outlined in Sect.\ \ref{sec:sample}, these galaxies show Balmer Decrements consistent with Case B recombination, indicative of negligible amounts of nebular attenuation.}}

The flux ratio between the gas-flow (\emph{broad}) and the static components (\emph{narrow}) lays in the range $F_{\rm B}/F_{\rm N} = 0.17-0.98$. {\ASL These values are typical of star-forming galaxies in the nearby Universe \citep[see the recent][]{Arribas2014, ReichardtChu2022, Xu2022-outflows}. A positive correlation between $F_{\rm B}/F_{\rm N}$ and stellar mass, extending below $10^8\msun$, has been suggested by some authors \citep[e.g.,][]{Newman2012, Freeman2019, Xu2025-outflows}, as the results of the stronger outflows powered by more massive galaxies. Some other works, however, do not find such trend \citep[][]{Swinbank2019, Perrotta2021, Concas2022}.} Interestingly, three out of five galaxies show an asymmetry in the [\oIII] line profile where the red wing is more prominent than the blue wing. {\ASL In other words, their gas-flow component is redshifted with respect to the static component, with a velocity offset $\Delta v = v_{\rm narrow} - v_{\rm broad} < 0$, beyond the $3\sigma$ significance in all cases.} This result is in contrast to the well-established picture of ionized outflows in the local Universe \citep[see][for a review]{ThompsonHeckman2024}, where blue-wing asymmetries are almost ubiquitous and attributed to gas-flows, since neither rotation nor mergers can account for such observed profile {\ASL \citep[i.e., they would not make the second component systematically broad and blue-shifted, see][]{Harrison2014}.} This being said, the presence of redshifted gas-flow components is not unseen \citep[e.g.,][]{RuschelDutra2021}. {\ASL Authors usually assume that the broad component primarily traces outflowing material, and the shift (blue or red) depends on the geometry and on the line of sight \citep[see discussion in][]{Llerena2023}. In the remaining of this section, we will follow this interpretation. Based on the low dust content of our galaxies, in Section \ref{sub:inflows} we will explore whether these flows can be explained as inflows on the near-side of the galaxy. However we are not able to definitively determine whether the flows are inflowing, and we conclude that outflows are the most likely origin of these flows. We will sometimes refer to both inflow and outflow candidates as \emph{gas flows} due to this uncertainty.}

{\ASL Following commonly used techniques in the literature,} we now proceed to characterize the physical properties of the outflow (see appendix's Table \ref{tab:data_gasflow}), namely the maximal outflow velocity (or \vflow), mass outflow rate (\mflow) and mass loading factor ($\eta$). We also perform a thorough literature comparison, by putting together the following data sets (sorted by redshift):
\begin{itemize}
    \item[$\circ$] {\citet{Marasco2023}}: sample of 19 nearby ($z \simeq 0$) star-forming galaxies with available optical VLT/MUSE spectroscopy. See the paper for references. 
    \item[$\circ$] {\citet{Llerena2023}}: selection of 35 star-forming galaxies at $z \simeq 3$ from the VUDS \citep{LeFevre2015} and VANDELS \citep{Garilli2021} surveys, with ground-based NIR spectroscopy from Keck/MOSFIRE \citep{Kriek2015, Cullen2021} and VLT/XShooter \citep{Amorin2017}.
    \item[$\circ$] {\citet{Carniani2024}}: high-resolution observations of 52 galaxies with NIRSpec G395H/F270LP at $z = 4-7$, as part of the JADES survey \citep{Eisenstein2023}.
    \item[$\circ$] {\citet{Xu2025}}: compilation of 130 {\it JWST} NIRSpec and NIRCam WFSS spectra at $z=3-9$, taken from various {\it JWST} programs (ERO, CEERS, FRESCO, GLASS, and JADES). See the paper for references. 
\end{itemize}

\noindent These samples have been carefully chosen to cover similar stellar mass ranges ($\log \mstar/\msun = 7-10$) but different look-back times (from local galaxies and Cosmic Noon, and beyond). Still, they have very distinct SFRs:  local dwarfs have low SFR overall \citep[${\rm SFR_{10} / \msun yr^{-1}} = 0.1-1$]{Marasco2023}, while  highly star-forming galaxies at Cosmic Noon have elevated SFRs \citep[${\rm SFR_{10} / \msun yr^{-1}} = 10-100$]{Llerena2023}. The high-redshift samples \citep{Carniani2024, Xu2025} show intermediate ${\rm SFR_{10} / \msun yr^{-1} = 1-10}$. 

\begin{figure*}
    \includegraphics[width=0.45\textwidth, page=1]{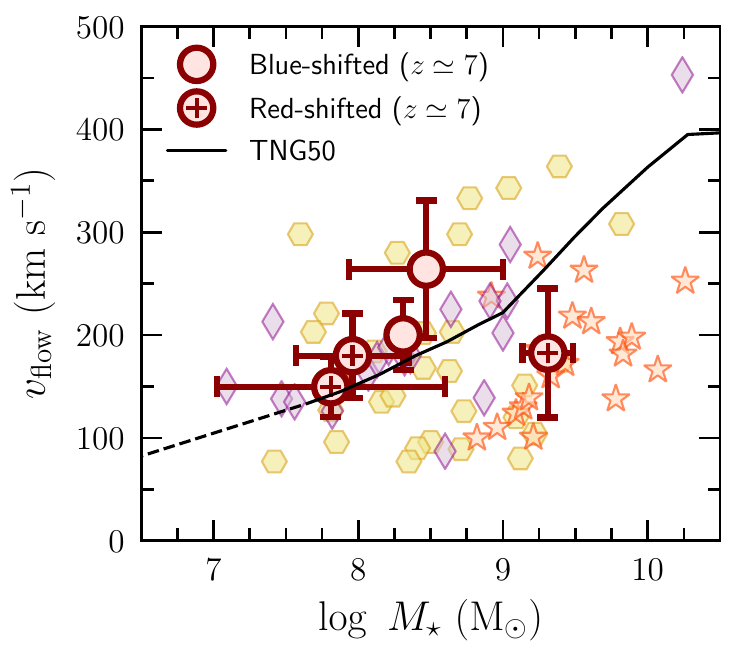}
    \includegraphics[width=0.45\textwidth, page=1]{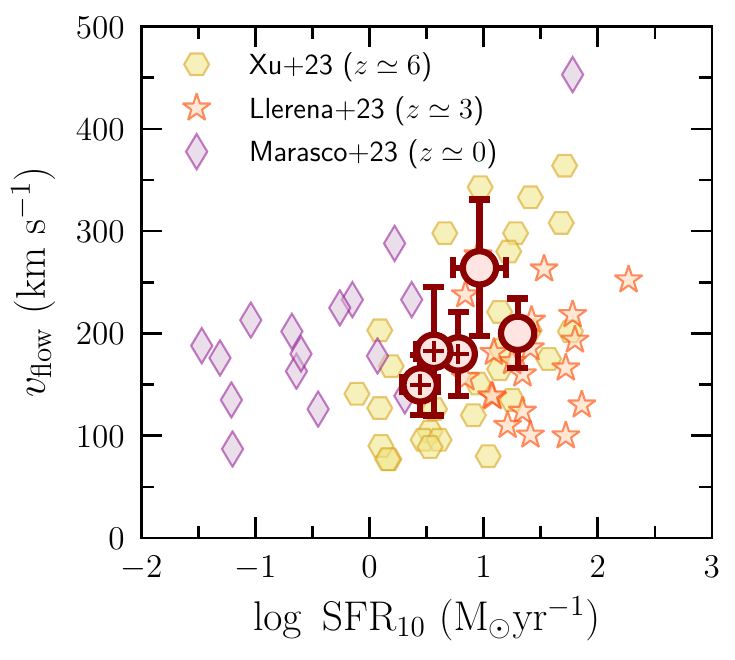}
    \includegraphics[width=0.45\textwidth, page=1]{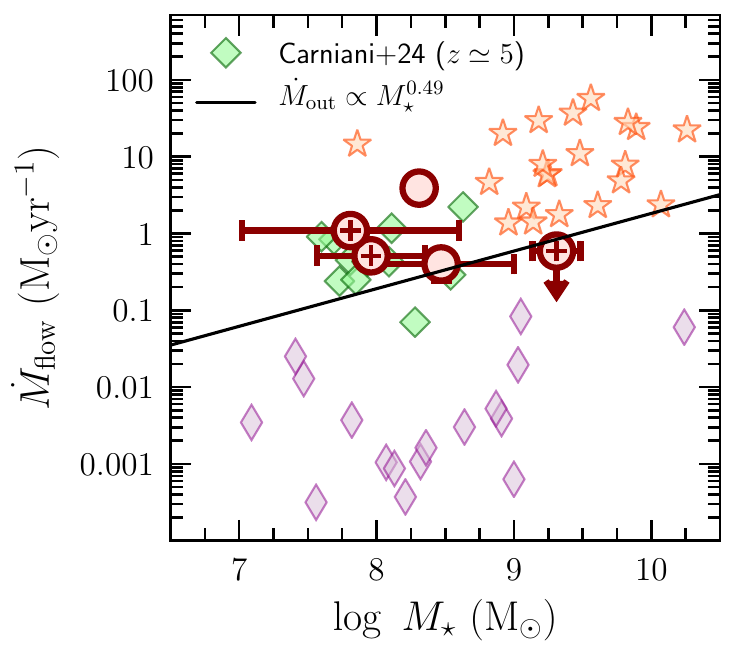}
    \includegraphics[width=0.45\textwidth, page=1]{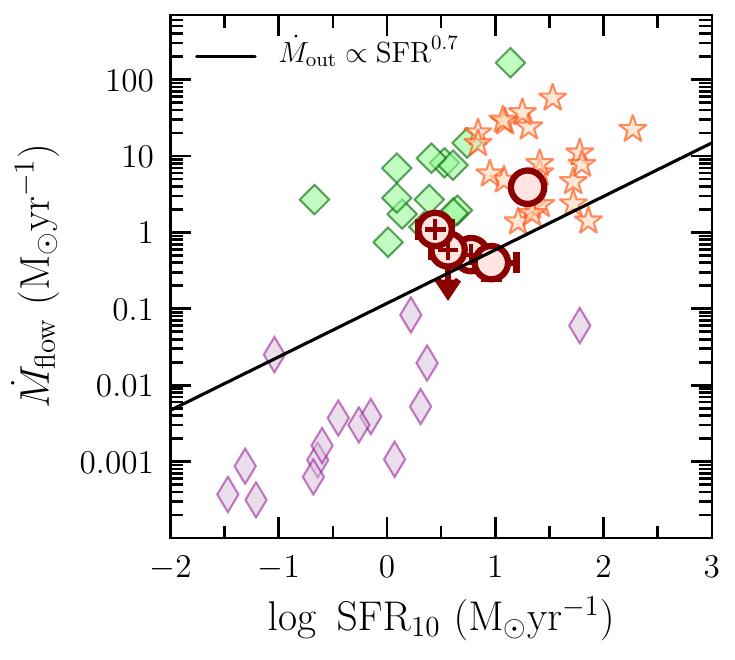}
\caption{{\bf Maximum gas-flow velocity (\vflow) and mass flow rate (\mflow) as a function of galaxy stellar mass (\mstar) and ${\rm SFR_{10}}$.} Thick circles show our high-$z$ gas-flows candidates {\ASL (galaxies showing red-shifted components are marked with a `$+$', the rest show blue-shifted components).} Literature measurements at $z \simeq 6$ \citep{Xu2025}, $z \simeq 5$ \citep{Carniani2024}, $z \simeq 3$ \citep{Llerena2023} and $z \simeq 0$ \citep{Marasco2023} are displayed via hexagons, thick diamonds, stars and thin diamonds, respectively. Black lines highlight the predictions from TNG50 cosmological simulations \citep[\emph{top-left:}][]{Nelson2019} and FIRE-like cosmological simulations \citep[\emph{bottom:}][]{Hopkins2012}.}
\label{fig:vout_mdot}
\end{figure*}

\subsection{{\ASL Outflow scaling relations}}
Without further ado, we first compute the maximum velocity of the outflow \citep[\vflow,][]{Rupke2005} as 
\begin{equation}
    \vflow = |v_{\rm broad} - v_{\rm narrow}| + \sigma_{\rm broad},
\end{equation}

\noindent where $\sigma_{\rm broad}$ is the velocity dispersion of the broad component of the [\oIII] line, and $v_i = c \cdot (\lambda_{i,0}-\lambda_{\rm rest})/\lambda_{\rm rest}$ corresponds to the velocity shift (in ${\rm km ~s^{-1}}$) between the narrow (static) and broad (outflow) components. The upper panel of Figure \ref{fig:vout_mdot} shows \vflow\ versus \mstar\ and SFR for our GO1871 gas-flow candidates, {\ASL where red-shifted components have been denoted with `$+$' and normal blue-shifted ones with empty symbols.} GO1871 outflows show moderate \vflow\ in the range ${\rm 150-250~km ~s^{-1}}$, compatible with other samples at high redshift \citep[e.g.,][]{Xu2025}. {\ASL The combination of all samples show a weak $\vflow-\mstar$ correlation with large scatter, only marginally consistent with the numerical prediction of an increasing \vflow\ with stellar mass \citep[e.g.,][]{Nelson2019}. The SFR averaged over the last 10~Myr (${\rm SFR_{10}}$) seems to be a more direct proxy of the speed of the outflow, where \vflow\ increasing with increasing ${\rm SFR_{10}}$ is more evident among the individual galaxy samples.} 

Next, we compute the mass of ionized gas that is being expelled by the outflow. {\ASL Assuming photo-ionization \citep{Carniani2015}, the gas mass of the ionized outflow ($M_{\rm flow}$) can be estimated from the luminosity of the [\oIII] broad component ($L^{\rm broad}_{[\oIII]}$) and the metallicity of the outflow as in,} 
\begin{equation}
    M_{\rm flow}{~(\msun)} = 0.6 \times 10^8 
    \left( \dfrac{L^{\rm broad}_{[\oIII]}}{10^{44}{\rm ~erg/s}}\right)~
    \left( \dfrac{Z_{\rm flow}}{Z_{\odot}} \right)^{-1}
    \left( \dfrac{n_{\rm flow}}{350 {\rm ~cm^{-3}}} \right)^{-1},
\end{equation}

\noindent where we assume $Z_{\rm flow}/Z_{\odot} = 0.1$ for the metallicity and $n_{\rm flow}/350{\rm ~cm^{-3}} = 1$ for the electron number density of the outflow. Ionized gas abundances of 10 per cent the solar value are consistent with the $R_{23}$ ratios measured in GO1871 galaxies (see Fig.\ \ref{fig:excitation}), and it is also typical of high-$z$ galaxies at these stellar masses \citep[see e.g.,][using the direct $T_e$ method]{Nakajima2023, Curti2024, Sanders2024}. Likewise, the chosen input value for the outflow density \citep[e.g.,][]{Isobe2023, Reddy2023} is based on independent measurements of the electron number density ($n_e$) using the [\ion{O}{ii}]$\lambda\lambda$3727,29 doublet (Stephenson et al., in preparation). 

\begin{figure}
    \includegraphics[width=0.95\columnwidth, page=1]{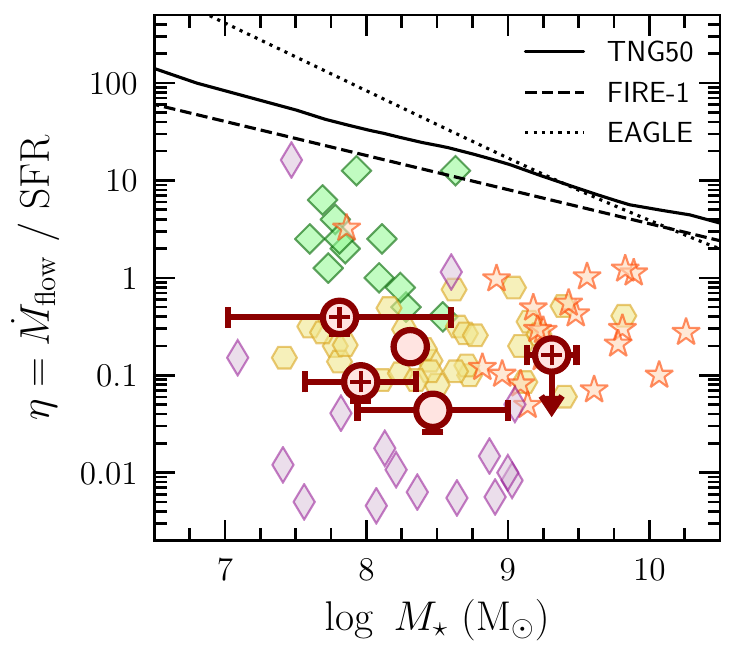}
    \caption{{\bf Mass loading factor ($\eta$) versus galaxy stellar mass (\mstar).} The solid, dashed and dotted lines show the median for different cosmological simulations: TNG50 \citep{Nelson2019} and EAGLE \citep{Mitchell2020}, both at $z = 2$, and FIRE-1 \citep[regardless of the redshift,][]{Muratov2015}, respectively. Symbols are the same as in Fig.\ \ref{fig:vout_mdot}. {\ASL At fixed \mstar, outflows in high-$z$ galaxies are \emph{more efficient} at removing gas from the star forming regions than in the local Universe}.}
\label{fig:eta_mstar}
\end{figure}

\begin{figure}
    \includegraphics[width=0.95\columnwidth, page=1]{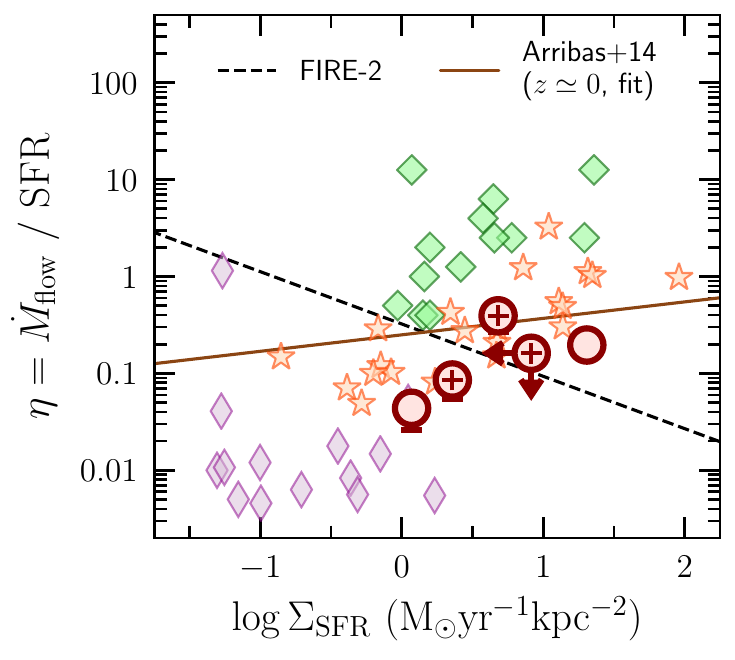}
    \caption{{\ASL {\bf The relation between the mass loading factor ($\eta$) and the SFR surface density (\sigSFR).} The red-solid and dashed lines correspond to the linear fits performed over the local observations from \citet{Arribas2014}, and to mock galaxies in the FIRE-2 simulations \citet{Pandya2021}. Symbols are the same as in Fig.\ \ref{fig:vout_mdot}. Galaxies that host a more compact star formation activity show more efficient outflows.}}
\label{fig:eta_sigmaSFR}
\end{figure}

Then, the mass outflow rate (\mflow) is defined as the amount of gas expelled due to the galactic wind per unit time. Assuming a spherical geometry, a unity covering fraction, and a constant \mflow\ with time \citep{Lutz2020}, this quantity can be computed as:
\begin{equation}
    \mflow = M_{\rm flow} \dfrac{\vflow}{r_{\rm flow}},
\end{equation}

\noindent where \vflow\ and $M_{\rm flow}$ are the maximal velocity of the outflow and the outflow mass calculated above. $r_{\rm flow}$ characterizes the extension of the outflow, and it is assumed to be twice the effective radius {\ASL ($2r_e$, i.e., the radius that encompasses \emph{all} the galaxy light)} measured in the NIRCam mosaics. The resulting mass outflow rates for GO1871 and literature samples are presented in the lower panel of Fig.\ \ref{fig:vout_mdot} as a function of stellar mass and ${\rm SFR_{10}}$. The dynamical range of the \mflow\ is large, with outflow rates in the nearby dwarfs of \citet{Marasco2023} falling four orders of magnitude apart from the sample of highly star-forming galaxies of \citet{Llerena2023} at the peak of cosmic star-formation \citep[see also][]{FS2019}. High-$z$ samples lay in between these two regimes, with moderate $\mflow / \msun {\rm yr^{-1}} = 0.2-5$ \citep[e.g.,][]{Carniani2024}. 

At fixed \mstar, literature \mflow\ measurements at different cosmic epochs span a wide range. An underlying \mflow\ correlation becomes more clear with SFR. Moving from low to high ${\rm SFR_{10}}$, mass outflow rates monotonically increase, with a slope that qualitatively matches the predictions from mock winds ($\mflow \propto {\rm SFR}^{0.7}$) in cosmological simulations including realistic feedback recipes \citep[FIRE-like simulation, see][]{Hopkins2012}. Measurements of \mflow\ in high-$z$ galaxies such as this work, \citet{Carniani2024} or \citet{Xu2025}, bridge the gap in the outflow landscape by probing moderate SFR and \mflow. {\arf Standing alone, however, our data set --spanning an order of magnitude in SFR-- does not show an evident correlation between the mass-outflow rate and SFR.}

Both moderate outflow velocities (a few hundreds ${\rm km~s^{-1}}$) and mass outflow rates comparable to the SFR are standard among starburst-driven winds in typical $L^{\star}$ galaxies in the local Universe \citep[e.g., see review article by][]{Veilleux2020}. Correlations between \vflow, \mflow, \mstar\ and SFR are expected for this type of winds \citep[e.g.,][]{Murray2005, Rupke2005, Heckman2015}. On the one hand, faster outflows (high \vflow) are usually associated with higher SFRs \citep{Arribas2014, Chisholm2015}: more vigorous star formation events would inject greater amounts of energy and momentum into the ISM, driving more powerful winds. Similarly, \mflow\ scales exponentially with SFR \citep{Xu2022-outflows, Marasco2023}, since the star-formation-driven feedback is proportional to the stellar production rate. On the other hand, the scaling with \mstar\ arises from the interplay between feedback energy and gravitational binding energy, introducing more scatter into the scaling relations {\ASL with stellar mass.}

{\ASL The efficiency of galactic outflows to regulate star formation is parameterized by the mass loading factor, $\eta$ \citep{Heckman1990}, measuring the amount of ejected gas relative to the rate at which the galaxy forms stars.} In other words, $\eta$ can be defined as the ratio of the mass outflow rate to the galaxy SFR \citep[e.g.,][]{Veilleux2005},
\begin{equation}
    \eta = \dfrac{\mflow}{\rm SFR},
\end{equation}

\noindent where for SFR we use ${\rm SFR_{10}}$. $\eta \geq 1$ can lead to gas depletion in the galaxy, reducing the potential for future star formation, {\ASL although the presence of gas in hotter phases, and the mass transfer between them makes this interpretation rather simplistic \citep[e.g.,][]{FieldingBrian2022, Steinwandel2024}.} 

Figure \ref{fig:eta_mstar} shows the mass loading factor for both our GO1871 ($\eta \simeq 0.04-0.4$) and literature samples, as a function of the galaxy stellar mass. Low-mass galaxies at high redshift \citep[][and this work]{Xu2025} show mass loading factors below unity. Again, observed mass loadings span three order of magnitudes, with some local dwarfs \citep{Marasco2023} having the least efficient winds. Cosmic Noon \citep{Llerena2023} and star-forming systems at higher redshifts show similar mass loadings, with the exception of \citet{Carniani2024}, exceeding beyond $\eta \geq 1$. {\ASL In conclusion, at fixed \mstar\ or SFR, outflows in high-$z$ galaxies are \emph{more efficient} in removing baryons than galaxies in the local Universe, in contrast with some of the proposed \emph{feedback-free} scenarios to explain the abundance of early galaxies \citep[e.g.,][]{Li2024, Dekel2023, Wang2023}. As we will argue in the next section, and thanks to their low dynamical masses, outflows in high-$z$ galaxies may have the ability to escape the gravitational potential of the host and enrich the CGM.}

According to theoretical expectations \citep{Murray2005} and numerical simulations \citep{Muratov2015}, {\arf the mass loading factor of star-formation driven winds should decrease with increasing stellar mass up to $\log \mstar/\msun \simeq 10$,} almost independently of redshift \citep{Nelson2019}. This means that the feedback in low-mass galaxies is more effective at expelling gas compared to more massive galaxies, where outflows may be suppressed by stronger gravitational potentials. The slope derived from simulations, $\eta \propto \mstar^{-0.3}$, is consistent with some works \citep[e.g.,][]{Chisholm2017, FS2019, Stanton2024}, although this conclusion becomes scarce once the different galaxy samples are looked at separately \citep[see also][]{McQuinn2019}. Perhaps more surprising is the fact that, {\ASL compared to the compilation of measurements in Fig.\ \ref{fig:eta_mstar},} cosmological simulations overshoot the normalization in the $\eta-\mstar$ relation, e.g., by more than a factor of $\times 10$ in the latest, state-of-the-art FIRE-2 simulation \citep{Pandya2021}. {\ASL This may indicate that there is a significant fraction of outflow mass in the hot phase, that the rest-optical nebular lines are unable to probe.} {\arf At masses higher than $\log \mstar/\msun \geq 10$, a possible turnover of the mass loading has been associated with AGN-driven winds by some observational \citep{FS2019, Swinbank2019, Concas2022} and numerical works \citep{Mitchell2020},} although this topic is constantly under revision \citep[see the recent][]{Weldon2024}. 

Finally, our gas-flow candidates also show some of the lowest $\mstar/\mdyn$ but highest \sigSFR\ within the GO1871 sample, with evidence of recent burst of star-formation in their SED via ${\rm SFR_{10}/SFR_{100} \geq 1}$, and whose intensity also scales with \sigSFR\ (Fig.\ \ref{fig:sample}). This reveals a picture in which high gas densities are needed in order to foster intense and compact star-forming events, that will eventually lead to the launch of star-formation driven outflows \citep[e.g.,][]{Heckman2002}. {\ASL In Figure \ref{fig:eta_sigmaSFR}, we investigate the relationship between the estimated mass loading factor and the compactness of the star formation (\sigSFR) for our high-$z$ galaxies and comparison samples at lower redshifts. Observationally, a positive, significant correlation exist between $\eta$ and \sigSFR\ \citep[see e.g., the linear fit by][]{Arribas2014}.} Consistently, intense outflows have been observed ubiquitously among high \sigSFR\ galaxies in other samples, showing faster velocities, higher mass outflow rates and increasing mass loading factors with \sigSFR\ \citep[e.g.,][]{Newman2012, Arribas2014, Davies2019, ReichardtChu2022, Llerena2023}. {\ASL In contrast, hydrodynamical simulations predict a negative trend between $\eta$ and \sigSFR, like due to the inverse proportionality of the $\eta-\Sigma_{\rm gas}$ relation \citep[e.g.,][]{Hopkins2012, Pandya2021}.}

\subsection{{\ASL Outflows enrich the CGM at high-$z$}}
\begin{figure}
    \includegraphics[width=0.95\columnwidth, page=1]{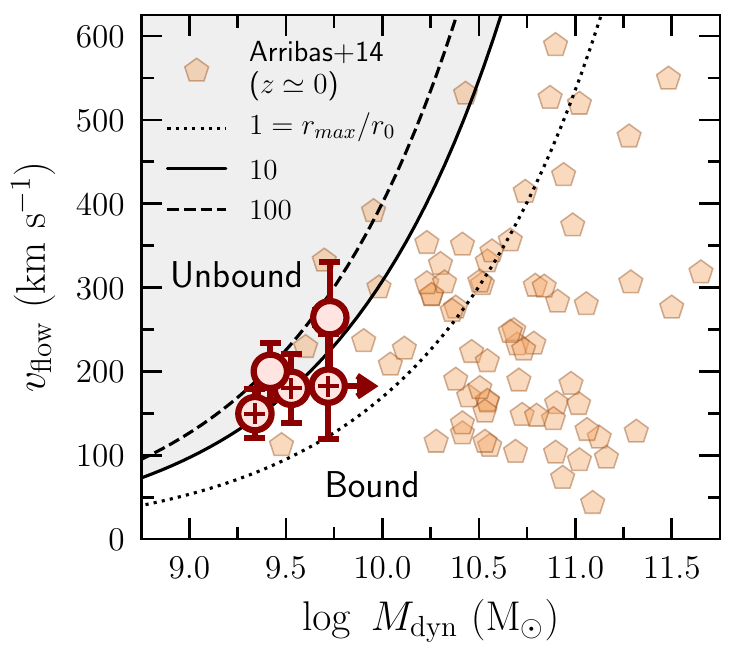}
\caption{{\bf Relation between the gas-flow velocity (\vflow) and the dynamical mass (\mdyn).} The local measurements by \citet{Arribas2014} are plotted with pentagons. The dotted, solid and dashed lines show the theoretical escape velocity ($v_{\rm esc}$) at $r_0 = 3$~kpc, as a function of the dynamical mass for gravitational potentials extending out to $r_{\rm max}/r_0 = 1, 10$ (default) and 100. Due to their low dynamical masses, galactic winds in low-mass galaxies at high-$z$ will likely escape the gravitational potential of the host galaxy {\ASL (\emph{unbound} region)}, enriching the surrounding CGM.}
\label{fig:vout_mdyn}
\end{figure}

An important question to address is whether the outflows detected in our sample of high-$z$ galaxies would remain bound to the gravitational influence of the host galaxy. In the case of metal-enriched outflows \citep[e.g.,][]{Chisholm2018, HamelBravo2024}, this will determine the ability of the gas to escape and reach the CGM \citep[e.g.,][]{Muratov2017}. To approach this question, we compare the outflow velocity to the escape velocity given by the dynamical mass of the host galaxy. An approximation to the mean escape velocity ($v_{\rm esc}$), {\ASL measured at a distance $r_0$,} is given in \citet{Arribas2014}, 
\begin{equation}
    v_{\rm esc} \approx \left( 
    \dfrac{2M_{\rm dyn}G \times (1+r_{\rm max}/r_0)}{3r_0} \right),
\end{equation}

\noindent assuming an isothermal gravitational potential that extends to a maximum radius $r_{\rm max}$ \citep{Heckman2000, Bellocchi2013}. Figure \ref{fig:vout_mdyn} shows our GO1871 measurements of \vflow\ versus \mdyn, together with the local comparison sample of \citet{Arribas2014}, at higher \mdyn's. The dotted, solid and dashed lines correspond to $r_{\rm max}/r_0 = 1, 10$ and 100 of the $v_{\rm esc}$ approximation, where the outflow velocity is measured at $r_0=3$~kpc (twice the median effective radius in our sample, approximately). We take the $r_{\rm max}/r_0 = 10$ curve as reference \citep[see e.g.,][]{Flury2023}. {\ASL As a result, points falling above the solid line in Fig. \ref{fig:vout_mdyn} will have outflow velocities higher than the escape velocity (\emph{unbound} region). While the blue-shifted outflow candidates in our sample fulfill this condition, those identified as red-shifted gas-flow candidates are consistent with below the curve (although still within 1$\sigma$). In the former case, part of the ejected gas could escape the gravitational potential of the host galaxy and contribute to enriching the CGM with metals \citep[e.g.,][]{Tumlinson2017, Yucheng2020}, in line with the conclusions outlined in \citet{Carniani2024}.} 

Last but not least, we highlight how the bulk of the Arribas et al. sample, composed of local luminous galaxies at $\log \mdyn/\msun \geq 10$, will most likely retain all their gas reservoirs: outflows will eventually cool down and fall back into the galaxy well in the form of galactic fountains. \citet{Xu2025} made a similar claim at high-$z$ that their reported velocities were not high enough for the outflow to escape from the galactic potentials, despite probing almost identical values of \mstar, SFR, \vflow\ and \mflow\ as in this work. Without measurement of \mdyn, Xu et al. {\ASL relied on additional assumptions for the escape velocity (i.e., they were computed from the theoretical circular velocity of the dark-matter halo),} this may be the origin of the discrepancy. {\ASL To sum up, our observation of outflows in low-mass, high-$z$ galaxies --with moderate velocities and mass lo ading factors-- imply that stellar feedback is important for driving the baryon cycle at $z > 5$.}

\subsection{{\ASL Tentative evidence of inflows in high-$z$ galaxies}}\label{sub:inflows}
{\ASL As mentioned above, three out of our five gas-flow candidates show red-shifted broad emission features in the [\ion{O}{iii}] profiles (Fig.\ \ref{fig:spectra}), with no blue-shifted components detected. In the literature, this is usually attributed to outflows in the far-side of the line of sight \citep{Arribas2014, Llerena2023}, and throughout the previous section we have also adopted this interpretation. However, given the negligible dust attenuation measured in these galaxies {\arf (Sect.\ \ref{sec:sample}),} one should expect to observe the blue-shifted counterpart of the expanding gas in the spectra as well \citep[e.g.,][]{RuschelDutra2021}. Instead, and invoking analogous geometrical arguments, it is worth considering the prospect that the red-shifted broad emission line traces inflows of ionized gas from the near-side of the galaxy. If true, the high detection rate of inflows in high-$z$ galaxies lays in contrast with observations of the nearby Universe, where the incidence of inflows is very low \citep[$\leq5\%$, from][]{Rubin2014}. This behavior is expected, as the specific gas accretion rate on to dark matter haloes decreases with decreasing redshift \citep[e.g.,][]{vandeVoort2011}, but it has not been tested observationally to date. 

The evidence of galactic inflows in our high-$z$ galaxies is supported by additional factors. {\arf For instance, the maximum velocity (Fig.\ \ref{fig:vout_mdot}) of the red-shifted components ($\vflow \leq 200{\rm km~s^{-1}}$) is lower than for the outflow candidates. Narrower spectroscopic features are expected to trace the inflowing gas, because the gas motions are collimated and less turbulent than those observed in the outflows \citep[e.g.,][]{Weldon2023}.} Indeed, the inferred escape velocity from the dynamical mass of the redshifted gas flows (Fig.\ \ref{fig:vout_mdyn}) is \emph{comparable} to the maximum velocity of inflowing gas, and marginally consistent with them being gravitationally bound to the host galaxy and accreting at roughly the virial velocity of the halo \citep[e.g.,][]{Goerdt2015}. Although an increase in velocity and mass inflow rate with \mstar\ and SFR is still expected, the \vflow\ and \mflow\ scaling relation with galaxy properties will be different than those for the outflows \citep[e.g.,][]{Martin2012, Roberts-Borsani2020}. Unfortunately, our limited sample size prevent us to draw further conclusions. 

Further, the mass loading factors are all below unity, which would indicate that the inflows are not sufficient to continue feeding the star-formation of the galaxy \citep[e.g.,][]{SAlmeida2014-review}. This hints that these galaxies will have a bursty SFH \citep[e.g.,][]{Muratov2015}, which is supported by the high burstiness parameters inferred from the SED, ${\rm SFR_{10}/SFR_{100} \geq 1}$. Finally, the high \sigSFR\ in these galaxies, and therefore the higher implied $\Sigma_{\rm gas}$, likely requires a high gas accretion rate and likely leads to chemical inhomogeneities \citep[e.g.,][]{SAlmeida2014, Cameron2021, GdVE2023}. In other words, gas accretion from the cosmic web can fuel the star-formation (boosting \sigSFR), increasing the gas fractions that would inevitably decrease the $\mstar/\mdyn$ ratios seen in these high-$z$ galaxies (Fig.\ \ref{fig:mstar_mdyn}). While tantalizing and a possible departure from local galaxies, the outflow scenario cannot be ruled out with the current data. Definitively testing the inflow hypothesis requires further observations \citep[e.g., absorption lines, see][]{Rubin2012}. Consistently, we have adopted the outflows emerging from the far-side of the ISM as the main physical picture of the gas flows detected in this paper.} 

\section{Conclusions}\label{sec:conclusions}
In this paper, we explore the dynamical mass and gas-flow properties of high-$z$ galaxies. To do so, we make use of high-resolution {\it JWST}/NIRSpec spectroscopy in the G235H/F170LP and G395H/F290LP gratings \citep{Jacobsen2022}. Our sample is composed of 16 moderately faint ($M_{\rm UV} \simeq -20$) star-forming galaxies at redshifts $4 \leq z \leq 7.6$ from program ID GO1871 (PI: Chisholm), ranging in $\log$ stellar masses of $\log \mstar/\msun = 7.75-9.75$, ${\rm SFR_{H\beta}}{/\msun {\rm yr}^{-1}} = 0.3 - 30$ and $\sigSFR /\msun {\rm yr^{-1} kpc^{-2}} = 0.3-20$. We summarize our main conclusions below. 

\begin{itemize}
    \item[$-$] The nebular line ratios reveal a metal-poor, dustless and highly ionized ISM, with $\log O_{32}=0.5-2$ and $\log R_{23}=0.5-1.5$ comparable to other star-forming samples at high-$z$ \citep[e.g.,][]{Cameron2023, Mascia2023, Sanders2023}, as well as to EELGs in the nearby Universe \citep[e.g.,][]{Yang2017, Izotov2020, Flury2022}. 
    \item[$-$] The width of emission lines such as H$\beta$ and [\oIII] is spectroscopically resolved. {\arf Accounting for instrumental broadening only, we report upper limits on the velocity dispersion of the ionized gas of $\sigma_{\rm gas}{\rm ~(km/s)} = 38-96$ in our high-$z$ sample, comparable to massive galaxies at lower redshifts \citep[e.g.,][]{FS2018}. {\ASL We argue that this decrease in velocity dispersion is due to the low stellar mass in these high-$z$ systems \citep[e.g.,][]{Pillepich2019}, even though the overall $\sigma_{\rm gas}$ increases at high redshift \citep[e.g.,][]{Ubler2019}}.}
    \item[$-$] Together with the galaxy sizes ($r_e = 400-960~$pc) modeled from archival NIRCam imaging \citep{Oesch23}, the resolved [\oIII] lines allow us to estimate the dynamical mass. Following \citet{Maiolino2024}, we obtain $\log \mdyn/\msun = 9.25-10.25$, which leads to surprisingly low stellar-to-dynamical mass ratios in these high-$z$ galaxies, i.e., $\log \mstar/\mdyn \in [-0.5,-2]$. 
    \item[$-$] The $\mstar/\mdyn$ ratio shows a tentative, decreasing correlation with increasing \sigSFR, suggesting higher gas-mass densities and gas fractions (average $f_{\rm gas} = 0.65$) with increasing $\mstar/\mdyn$ ratio. As widely discussed in \citet{deGraaff2024}, the contribution of the inferred gas masses to the total baryonic mass does not solve the mass discrepancy. To reconcile the mass ratios, $M_{\rm baryon}/\mdyn = (\mstar + M_{\rm gas})/\mdyn \simeq 0.2-0.7$, would require invoking other mechanisms such as a significant decrease of the star-formation efficiency, {\ASL as opposed to recently proposed scenarios to explain the over-abundance of UV-bright galaxies at high-redshift with {\it JWST} \citep[e.g.,][]{Wang2023}.} 
    \item[$-$] Evidence of gas-flows is found in five out of 16 galaxies ($\simeq 30$ per cent incidence rate) based upon the statistical need of  broad components to reproduce the wings of the [\oIII] lines. We compute the properties of the gas flow, namely outflow velocities ($\vflow{\rm ~(km/s)} = 150-250$) and mass outflow rates ($\mflow{\rm /\msun yr^{-1}} = 0.2-5$), reporting values comparable to \citet{Xu2025} at similarly high redshifts. 
    \item[$-$] In the context of starburst-driven outflows, low-mass, high-$z$ galaxies bridge the gap in the outflow scaling relations between the dwarf regime in the nearby Universe \citep{Marasco2023}, and more massive, highly star-forming galaxies at Cosmic Noon \citep{Llerena2023}, showing moderate \vflow\ and \mflow, and mass loading factors $\eta = 0.04-0.4$. {\ASL As a result, warm ionized outflows in high-$z$ galaxies are \emph{more efficient} in removing baryons from star-forming regions than galaxies in the local Universe with the same stellar mass.} 
    \item[$-$] Our high-$z$ outflow candidates are also among the lowest $\mstar/\mdyn$ and highest \sigSFR\ in the sample, showing that more recent bursts of star-formation (${\rm SFR_{10}/SFR_{100} \geq 1}$) have elevated star formation rate surface densities (\sigSFR). These findings emphasize the need for both high gas-mass surface densities (high \sigSFR) and bursty SFHs to foster intense, compact star-formation events that launch galactic winds \citep[e.g.,][]{Heckman2002, ReichardtChu2022}.  
    \item[$-$] We compare the outflow velocity to the escape velocity given by the dynamical mass of the galaxy \citep[following][]{Arribas2014}, and find that these high-$z$ ionized outflows will likely escape the gravitational influence of the host, potentially enriching the CGM with metals \citep[see also][]{Carniani2024}. 
    \item[$-$] {\ASL Three out of the five gas-flow candidates show tentative signatures of inflows, with broad, red-shifted components in the [\oIII] line profile. In this cases, the maximum velocity of the inflow is lower than the outflow velocity in the other candidates, and below the escape velocity that is estimated from the dynamical mass. If confirmed, our findings would imply a much higher incidence of inflows compared to the nearby Universe \citep[e.g.,][]{Martin2012, Rubin2014}. The accretion of gas would also increase the gas fractions, in agreement with the high \sigSFR, low $\mstar/\mdyn$ ratios, and bursty SFHs in these high-$z$ systems.}
\end{itemize}

{\ASL This work highlights the importance of high-resolution spectroscopic surveys with NIRSpec to study the baryon cycle of galaxies beyond Cosmic Noon. Providing one of the first evidences of stellar feedback and baryonic budget during the Epoch of Reionization, our results imply that inflows and outflows are important for driving the baryon cycle at early epochs of galaxy formation.}

\section*{Acknowledgements}
% The authors thank the anonymous referee for providing useful comments, which have certainly improved the quality of this paper. 
The authors thank M. G. del Valle-Espinosa and T. M. Stanton for helping with figure formatting. ASL acknowledges support from Knut and Alice Wallenberg Foundation. MJH is supported by the Swedish Research Council (Vetenskapsrådet) and is Fellow of the Knut and Alice Wallenberg Foundation. NGG and YII acknowledge support from the National Academy of Sciences of Ukraine by its project no.0123U102248 and from the Simons Foundation. This work is based (in part) on observations made with the NASA/ESA/CSA James Webb Space Telescope. The data were obtained from the Mikulski Archive for Space Telescopes at the Space Telescope Science Institute, which is operated by the Association of Universities for Research in Astronomy, Inc., under NASA contract NAS 5-03127 for JWST. These observations are associated with program \#01871.

%%%%%%%%%%%%%%%%%%%%%%%%%%%%%%%%%%%%%%%%%%%%%%%%%%

%%%%%%%%%%%%%%%%%%%% REFERENCES %%%%%%%%%%%%%%%%%%

% The best way to enter references is to use BibTeX:

\section*{Data availability}
All JWST data used in this paper can be found in MAST, under DOI: \href{https://archive.stsci.edu/doi/resolve/resolve.html?doi=10.17909/vchg-y922}{10.17909/vchg-y922}. Any additional product and/or data underlying this article will be shared on reasonable request to the corresponding author.

\bibliographystyle{mnras}
\bibliography{references_outflows}% if your bibtex file is called example.bib

% Alternatively you could enter them by hand, like this:
% This method is tedious and prone to error if you have lots of references
%\begin{thebibliography}{99}
%\bibitem[\protect\citeauthoryear{Author}{2012}]{Author2012}
%Author A.~N., 2013, Journal of Improbable Astronomy, 1, 1
%\bibitem[\protect\citeauthoryear{Others}{2013}]{Others2013}
%Others S., 2012, Journal of Interesting Stuff, 17, 198
%\end{thebibliography}

%%%%%%%%%%%%%%%%%%%%%%%%%%%%%%%%%%%%%%%%%%%%%%%%%%

%%%%%%%%%%%%%%%%% APPENDICES %%%%%%%%%%%%%%%%%%%%%

\appendix
\section{Data tables}\label{app:data_tables}
In this appendix, we compile the main data and products used in this article. Specifically, Table \ref{tab:data_sample} lists the physical and emission line properties of GO1871 galaxies; Table \ref{tab:data_morphology} summarizes the results from the morphological modeling and subsequent dynamical quantities; and Table \ref{tab:data_gasflow} includes the gas-flow derived parameters for the gas flow candidates in the GO1871 sample. 

\renewcommand{\arraystretch}{1.25}
\begin{table*}
\begin{center}
\caption{Physical properties derived from SED fitting and emission line ratios for GO1871 galaxies.}
\begin{tabular}{cccccccccc}
\toprule
MSAid & RA & Dec & $z_{\rm spec}$ & $M_{\rm UV}$ & $\log M_{\star}$ & ${\rm SFR_{10} }$ & ${\rm SFR_{100} }$ & $\log O_{32}$ & $\log R_{23}$ \\
& (deg.) & (deg.) & ${\rm (AB)}$ & ${\rm (\msun) }$ & ${\rm (\msun yr^{-1}) }$ & ${\rm (\msun yr^{-1}) }$ & & \\
\midrule
$1871-9$ & $189.16626246$ & $62.31651129$ & $6.5691$ & $-19.7$ & $9.00~\pm~1.17$ & $1.6_{-1.6}^{+3.3}$ & $2.3_{-1.6}^{+3.3}$ & $-$ & $-$ \\
$1871-10$ & $189.22199791$ & $62.31576868$ & $7.5990$ & $-20.0$ & $7.96~\pm~0.39$ & $6.0_{-1.6}^{+1.2}$ & $0.9_{-0.2}^{+0.8}$ & $\leq 1.04$ & $\geq 0.63$ \\
$1871-11$ & $189.22436984$ & $62.31137047$ & $7.6100$ & $-21.1$ & $9.07~\pm~0.31$ & $0.1_{-0.1}^{+0.7}$ & $9.7_{-2.5}^{+7.0}$ & $\leq 1.60$ & $\geq 1.07$ \\
$1871-12$ & $189.15797785$ & $62.30239809$ & $7.5020$ & $-21.1$ & $8.31~\pm~0.09$ & $19.9_{-2.1}^{+3.0}$ & $2.0_{-0.2}^{+0.3}$ & $0.73~\pm~0.01$ & $0.96~\pm~0.05$ \\
$1871-14$ & $189.22515469$ & $62.28628471$ & $7.2049$ & $-19.9$ & $9.31~\pm~0.17$ & $3.7_{-1.2}^{+17.8}$ & $3.2_{-1.6}^{+2.3}$ & $0.95~\pm~0.01$ & $1.08~\pm~0.08$ \\
$1871-29$ & $189.18882760$ & $62.26991420$ & $5.0191$ & $-20.3$ & $9.24~\pm~0.22$ & $0.0_{-0.0}^{+0.1}$ & $10.7_{-3.0}^{+3.1}$ & $0.64~\pm~0.01$ & $1.04~\pm~0.04$ \\
$1871-40$ & $189.21508726$ & $62.31794769$ & $7.0948$ & $-20.5$ & $8.47~\pm~0.53$ & $9.2_{-3.1}^{+4.8}$ & $2.2_{-0.9}^{+2.4}$ & $\leq 0.64$ & $\geq 0.69$ \\
$1871-63$ & $189.17219338$ & $62.30563914$ & $5.5872$ & $-19.9$ & $7.84~\pm~0.34$ & $4.9_{-0.9}^{+0.9}$ & $0.6_{-0.1}^{+0.3}$ & $1.54~\pm~0.00$ & $0.89~\pm~0.01$ \\
$1871-70$ & $189.20956983$ & $62.29928187$ & $7.0302$ & $-19.3$ & $8.85~\pm~0.41$ & $1.5_{-1.1}^{+1.1}$ & $1.7_{-0.9}^{+1.3}$ & $\leq 0.68$ & $\geq 0.56$ \\
$1871-105$ & $189.14873069$ & $62.28092990$ & $6.5669$ & $-19.5$ & $7.81~\pm~0.79$ & $2.8_{-1.2}^{+1.0}$ & $0.5_{-0.2}^{+1.6}$ & $-$ & $-$ \\
$1871-451$ & $189.23835414$ & $62.28442312$ & $4.4615$ & $-19.0$ & $7.99~\pm~0.59$ & $1.7_{-1.0}^{+0.6}$ & $0.6_{-0.3}^{+0.8}$ & $0.31~\pm~0.02$ & $0.81~\pm~0.07$ \\
$1871-545$ & $189.14071083$ & $62.27725108$ & $4.0416$ & $-19.3$ & $9.41~\pm~0.41$ & $66.7_{-39.7}^{+38.6}$ & $16.6_{-5.9}^{+9.6}$ & $-0.02~\pm~0.01$ & $0.86~\pm~0.02$ \\
$1871-912$ & $189.18613720$ & $62.27089253$ & $6.7161$ & $-21.8$ & $9.69~\pm~0.20$ & $3.8_{-3.6}^{+9.8}$ & $27.1_{-11.4}^{+14.8}$ & $0.67~\pm~0.01$ & $1.06~\pm~0.05$ \\
$1871-918$ & $189.20254167$ & $62.27558333$ & $6.9062$ & $-20.6$ & $8.78~\pm~0.58$ & $6.5_{-4.3}^{+5.7}$ & $3.9_{-2.4}^{+3.6}$ & $1.27~\pm~0.04$ & $0.62~\pm~0.12$ \\
$1871-969$ & $189.15554992$ & $62.28678755$ & $6.5613$ & $-19.0$ & $7.80~\pm~0.55$ & $1.5_{-1.2}^{+0.7}$ & $0.5_{-0.3}^{+0.7}$ & $-$ & $-$ \\
$1871-1032$ & $189.19565216$ & $62.28247165$ & $7.0889$ & $-19.7$ & $9.07~\pm~0.24$ & $0.0_{-0.0}^{+0.2}$ & $6.3_{-1.7}^{+1.3}$ & $\leq 0.69$ & $\geq 1.24$ \\
\bottomrule
\end{tabular}
\label{tab:data_sample}
\end{center}
{\bf Notes.} Column 1: MSA object identifier. {\ASL Column 2: source coordinates (RA and Dec).} Column 3: spectroscopic redshift. Column 4: UV magnitude (AB). Column 5: stellar mass (in \msun) from SED fitting. Columns 6 and 7: SED-derived SFRs (in $\msun {\rm yr}^{-1}$), averaged over 10Myr and 100Myr. Columns 8 and 9: $O_{32}$ (ionization) and $R_{23}$ (excitation) ratios. 
\end{table*}

\begin{table*}
\begin{center}
\caption{Morphological characterization and dynamical properties of GO1871 galaxies.}
\begin{tabular}{ccccccccccc}
\toprule
MSAid & ${\rm NIRCam}$ & ${\rm model}$ & $r_e$ & ${\rm q}$ & ${\rm PA}$ & $r_{\rm phys}$ & $\Sigma_{\rm SFR}$ & $\sigma_{\rm res}$ & $\sigma^*_{\rm gas}$ & $\log M_{\rm dyn}$ \\
& ${\rm band}$ & & ${\rm ~(arcsec)}$ & & ${\rm (deg.)}$ & ${\rm (pc)}$ & ${\rm (\msun yr^{-1} kpc^{-2})}$ & ${\rm (km~s^{-1})}$ & ${\rm (km~s^{-1})}$ & ${\rm (\msun) }$ \\
\midrule
$1871-9$ & ${\rm F182M}$ & PSF & $\leq 0.07$ & $0.00$ & $0$ & $\leq 404$ & $\geq 6.0$ & $30.1$ & $\leq 52.5$ & $\leq 9.64$ \\
$1871-10$ & ${\rm F444W}$ & S\'ersic & $0.14~\pm~0.05$ & $0.17$ & $68$ & $739~\pm~262$ & $2.3~\pm~0.3$ & $36.5$ & $65.2~\pm~5.3$ & $9.54~\pm~0.16$ \\
$1871-11$ & ${\rm F444W}$ & S\'ersic & $0.17~\pm~0.01$ & $0.23$ & $161$ & $846~\pm~37$ & $0.3~\pm~0.2$ & $29.4$ & $60.9~\pm~12.9$ & $9.77~\pm~0.15$ \\
$1871-12$ & ${\rm F444W}$ & S\'ersic & $0.07~\pm~0.01$ & $0.44$ & $114$ & $381~\pm~39$ & $20.3~\pm~2.1$ & $32.7$ & $67.3~\pm~2.8$ & $9.41~\pm~0.05$ \\
$1871-14$ & ${\rm F444W}$ & PSF & $\leq 0.07$ & $0.00$ & $0$ & $\leq 383$ & $\geq 8.2$ & $31.8$ & $\leq 60.5$ & $\leq 9.72$ \\
$1871-29$ & ${\rm F444W}$ & S\'ersic & $0.13~\pm~0.01$ & $0.60$ & $95$ & $839~\pm~83$ & $0.7~\pm~0.1$ & $38.5$ & $72.0~\pm~0.7$ & $9.96~\pm~0.04$ \\
$1871-40$ & ${\rm F182M}$ & S\'ersic & $0.14~\pm~0.01$ & $0.38$ & $177$ & $740~\pm~45$ & $1.2~\pm~0.3$ & $31.6$ & $60.6~\pm~5.9$ & $9.73~\pm~0.07$ \\
$1871-63$ & ${\rm F444W}$ & PSF & $\leq 0.07$ & $0.00$ & $0$ & $\leq 442$ & $\geq 9.1$ & $38.5$ & $\leq 45.0$ & $\leq 9.66$ \\
$1871-70$ & ${\rm F182M}$ & S\'ersic & $0.12~\pm~0.01$ & $0.43$ & $105$ & $659~\pm~52$ & $1.7~\pm~0.5$ & $29.0$ & $41.3~\pm~8.0$ & $9.42~\pm~0.12$ \\
$1871-105$ & ${\rm F182M}$ & S\'ersic & $0.07~\pm~0.01$ & $0.25$ & $34$ & $393~\pm~41$ & $4.8~\pm~0.4$ & $34.7$ & $49.9~\pm~1.4$ & $9.34~\pm~0.05$ \\
$1871-451$ & ${\rm F182M}$ & S\'ersic & $0.08~\pm~0.01$ & $0.67$ & $78$ & $556~\pm~66$ & $0.8~\pm~0.1$ & $26.7$ & $40.7~\pm~2.5$ & $9.37~\pm~0.06$ \\
$1871-545$ & ${\rm F444W}$ & S\'ersic & $0.12~\pm~0.01$ & $0.70$ & $57$ & $828~\pm~88$ & $1.2~\pm~0.1$ & $29.9$ & $96.4~\pm~1.4$ & $10.11~\pm~0.05$ \\
$1871-912$ & ${\rm F444W}$ & S\'ersic & $0.18~\pm~0.01$ & $0.46$ & $116$ & $970~\pm~59$ & $1.0~\pm~0.1$ & $34.7$ & $44.4~\pm~1.9$ & $9.68~\pm~0.04$ \\
$1871-918$ & ${\rm F444W}$ & PSF & $\leq 0.07$ & $0.00$ & $0$ & $\leq 393$ & $\geq 5.7$ & $29.4$ & $\leq 62.1$ & $\leq 9.73$ \\
$1871-969$ & ${\rm F210M}$ & PSF & $\leq 0.07$ & $0.00$ & $0$ & $\leq 405$ & $\geq 2.4$ & $31.7$ & $\leq 57.5$ & $\leq 9.71$ \\
$1871-1032$ & ${\rm F182M}$ & S\'ersic & $0.11~\pm~0.01$ & $0.42$ & $83$ & $600~\pm~37$ & $0.3~\pm~0.3$ & $36.1$ & $37.3~\pm~12.1$ & $9.41~\pm~0.16$ \\
\bottomrule
\end{tabular}
\label{tab:data_morphology}
\end{center}
{\bf Notes.} Column 1: MSA object identifier. Column 2 and 3: NIRCam band and model used in the morphological fitting. Columns 4, 5, 6 and 7: best-fit effective radius (in arcsec), axis ratio, position angle (in deg.) {\ASL and physical radius (in pc).} Column 8: SFR surface density (in ${\rm \msun yr^{-1} kpc^{-2}}$). Column 9: instrumental resolution from \textsc{MSAfit}. Column 10: velocity dispersion of the ionized gas, {\arf without accounting for the contributions of rotational components} (in ${\rm km~s^{-1}}$) as given by the width of the [\oIII]$\lambda$5007 lines, corrected by instrumental broadening. Column 11: inferred galaxy dynamical mass (in \msun). 
\end{table*}

\begin{table*}
\begin{center}
\caption{Gas flow parameters of GO1871 galaxies.}
\begin{tabular}{ccccccc}
\toprule
MSAid & ${\rm F_{\rm B}/F_{\rm N}}$ & $\sigma_{\rm narrow}$ & $\sigma_{\rm broad}$ & $v_{\rm flow}$ & $\dot{M}_{\rm flow}$ & $\eta$ \\
& & ${\rm (km~s^{-1})}$ & ${\rm (km~s^{-1})}$ & ${\rm (km~s^{-1})}$ & ${\rm (\msun yr^{-1})}$ & \\
\midrule
$1871-10$ & $0.98~\pm~0.32$ & $49.08~\pm~6.40$ & $121.70~\pm~29.95$ & $(-)~179.90~\pm~41.03$ & $0.51~\pm~0.16$ & $0.09~\pm~0.03$ \\
$1871-12$ & $0.84~\pm~0.13$ & $59.75~\pm~3.83$ & $160.20~\pm~19.22$ & $(+)~200.12~\pm~34.00$ & $3.90~\pm~0.78$ & $0.20~\pm~0.04$ \\
$1871-14$ & $0.17~\pm~0.05$ & $61.48~\pm~2.16$ & $110.64~\pm~56.12$ & $(-)~182.47~\pm~62.74$ & $\leq 0.59$ & $\leq 0.16$ \\
$1871-40$ & $0.81~\pm~0.28$ & $60.35~\pm~5.61$ & $259.15~\pm~60.62$ & $(+)~264.08~\pm~66.79$ & $0.40~\pm~0.16$ & $0.04~\pm~0.02$ \\
$1871-105$ & $0.42~\pm~0.10$ & $57.32~\pm~1.98$ & $103.17~\pm~8.73$ & $(-)~149.68~\pm~29.37$ & $1.09~\pm~0.31$ & $0.39~\pm~0.13$ \\
\bottomrule
\end{tabular}
\label{tab:data_gasflow}
\end{center}
{\bf Notes.} Column 1: MSA object identifier. Column 2: broad-to-narrow flux ratio. Columns 3 and 4: velocity dispersion of the narrow and broad components (in ${\rm km ~s^{-1}}$). Column 5: maximum gas-flow velocity (in ${\rm km ~s^{-1}}$). {\ASL The ($+$) symbols denote red-shifted broad components respect to the narrow profiles, while sources marked with ($-$) are blue-shifted components.} Column 6: mass flow rate (in ${\rm \msun yr^{-1}}$). Column 7: estimated mass loading factor. 
\end{table*}

%%%%%%%%%%%%%%%%%%%%%%%%%%%%%%%%%%%%%%%%%%%%%%%%%%

% Don't change these lines
\bsp	% typesetting comment
\label{lastpage}
\end{document}